%% file: cch-arxiv.tex
\documentclass[11pt]{article}
\usepackage{epsf,amsmath,amssymb,graphicx,scalefnt,hyperref}
\usepackage{rotating,ifthen,color,filemod}
\usepackage[noadjust]{cite}
\usepackage{bashful}
\usepackage{multirow}
\usepackage{dcolumn}
\newcommand{\abbrev}{\scalefont{.9}}
\newcolumntype{.}{D{.}{.}{2.14}}
\newcommand{\sushi}{{\tt SusHi}}

\newcommand{\pdfnlo}{{\tt MSTW2008nlo}}
\newcommand{\pdfnnlo}{{\tt MSTW2008nnlo}}
\newcommand{\citere}[1]{Ref.\,\cite{#1}}
\newcommand{\citeres}[1]{Refs.\,\cite{#1}}
\newcommand{\fs}[1]{#1{\abbrev FS}}
\newcommand{\qqh}{\Q'\bar \Q\phi}
\newcommand{\betaqq}{\beta_{\scriptscriptstyle QQ'}}
\newcommand{\bbH}{b\bar bH}
\newcommand{\bbh}{b\bar b\phi}
\newcommand{\bch}{b\bar c\phi^+}
\newcommand{\bsh}{b\bar s\phi}
\newcommand{\cch}{c\bar c\phi}
\newcommand{\csh}{c\bar s\phi^-}
\newcommand{\Q}{{\scriptstyle Q}}
\newcommand{\myA}{{\abbrev A}}
\newcommand{\myB}{{\abbrev B}}
\newcommand{\myC}{{\abbrev C}}

\newcommand{\eqn}[1]{Eq.\,(\ref{#1})}
\newcommand{\eqns}[1]{Eqs.\,(\ref{#1})}
\newcommand{\noeqn}[1]{(\ref{#1})}
\newcommand{\fig}[1]{Fig.\,\ref{#1}}
\newcommand{\figs}[1]{Figs.\,\ref{#1}}

\newcommand{\dd}{{\rm d}}

\newcommand{\lhc}{{\abbrev LHC}}
\newcommand{\qcd}{{\abbrev QCD}}
\newcommand{\sm}{{\abbrev SM}}

\newcommand{\bsm}{{\abbrev BSM}}
\newcommand{\pdf}{{\abbrev PDF}}
\newcommand{\lo}{{\abbrev LO}}
\newcommand{\nlo}{{\abbrev NLO}}
\newcommand{\nnlo}{{\abbrev NNLO}}
\newcommand{\nklo}[1]{{\abbrev N$^{#1}$LO}}

\newcommand{\muF}{\mu_\text{F}}
\newcommand{\muR}{\mu_\text{R}}
\newcommand{\muFhat}{\hat{\mu}_\text{F}}
\newcommand{\muRhat}{\hat{\mu}_\text{R}}
\newcommand{\mhiggs}{m_H}
\newcommand{\mphi}{m_\phi}

\newcommand{\mbottom}{m_b}
\newcommand{\mcharm}{m_c}
\title{%
\vspace*{-8em}{
\begin{flushright}
\mbox{\small\sf TTH-15-38, December 2015}\\[6em]
\end{flushright}
Higgs production in heavy quark annihilation through
next-to-next-to-leading order \qcd{}}}
\author{Robert V.~Harlander\\[1em]
{\it Institute for Theoretical Particle Physics and Cosmology}\\
{\it RWTH Aachen University, D-52056 Aachen, Germany}\\
{\tt robert.harlander@cern.ch}}
\date{}

\begin{document}
\maketitle
\begin{abstract}
The total inclusive cross section for charged and neutral Higgs
production in heavy-quark annihilation is presented through \nnlo{}
\qcd{}. It is shown that, aside from an overall factor, the partonic
cross section is independent of the initial-state quark flavors, and
that any interference terms involving two different Yukawa couplings
vanish. A simple criterion for defining the central renormalization and
factorization scale is proposed.  Its application to the $\bbh$ process
yields results which are compatible with the values usually adopted for
this process. Remarkably, we find little variation in these values for
the other initial-state quark flavors.  Finally, we disentangle the
impact of the different parton luminosities from genuine hard \nnlo{}
effects and find that, for the central scales, a naive rescaling by the
parton luminosities approximates the full result remarkably well.
\end{abstract}

\section{Introduction}

Models with an extended Higgs sector typically predict a spectrum of
Higgs bosons with very diverse properties (see, e.g.,
\citere{Gunion:1989we}). This means that the relative importance of
individual processes for the total production cross section can be very
different compared to the Standard Model (\sm{}) Higgs boson $H$, where
the main contribution to the total cross section at the Large Hadron
Collider (\lhc{}) is given by gluon fusion, $gg\to H$ (see, e.g.,
\citeres{Dittmaier:2011ti,Dittmaier:2012vm,Heinemeyer:2013tqa}). In
particular, quark-associated production can be much more important than
for \sm{} Higgs production.  For example, in supersymmetric theories, it
it can naturally occur that at least one of the neutral Higgs bosons
$\phi$ would be predominantly produced at the \lhc{} in association with
bottom-quarks, $pp\to \phi b\bar b$.  Also an enhanced coupling to charm
quarks can occur in many beyond-the-\sm{} (\bsm) scenarios, leading to
non-negligible contributions of associated Higgs-charm
production\,\cite{Delaunay:2013pja}. Similarly, the cross section for
charged Higgs bosons $\phi^\pm$ may receive contributions from
associated $c\bar s\phi^-/\bar cs\phi^+$ or $c\bar b\phi^-/\bar
cb\phi^+$ production, and we could even imagine flavor-violating
contributions of the form $(b\bar s+\bar bs)\phi$ to neutral Higgs
production\,\cite{Gomez:2015duj}.

The proper theoretical description of associated $\bbh$ production has a
long and still ongoing history. The main argument has been centered
around the question whether the so-called 4- or 5-flavor scheme
(referred to as \fs{4} or \fs{5} in what follows) is more approprate to
obtain the best approximation of the total inclusive cross section. In
the \fs{4}, bottom-quark parton densities are neglected, so that the
dominant leading-order (\lo{}) partonic process for $\bbh$ production is
$gg\to b\bar b\phi$ (the cross section for $q\bar q\to b\bar b\phi$ is
about a factor of ten smaller at the \lhc{}). Integration over the final
state bottom quark momenta leads to logarithms of the form
$\ln(\mbottom/\mphi)$ in the total inclusive Higgs production cross
section, where $\mbottom$ and $\mphi$ is the bottom-quark and the
Higgs-boson mass, respectively. The \fs{5} resums these terms to all
orders in the strong coupling $\alpha_s$ by introducing bottom-quark
parton densities, and describing the \lo{} partonic cross section as
$b\bar b\to \phi$. In the partonic calculation, the bottom-quark mass is
set to zero (except where it occurs in the Yukawa coupling), and all
collinear divergences are absorbed into the parton density functions
(\pdf{}s) through mass factorization. Concerning the sub-process $gg\to
\phi b\bar b$, there is a potential mismatch of this approach with the
treatment of the bottom-quark threshold in the parton
densities. However, by comparing the massless with a massive calculation
in this sub-channel\,\cite{Buttar:2006zd}, such effects could be shown
to be negligible w.r.t.\ the overall theoretical accuracy.

The current experimental analyses are based on a
combination of results from both approaches through the so-called
Santander-matching formula\,\cite{Harlander:2011aa}, where the \fs{4}
and \fs{5} results---the former at next-to-leading order
(\nlo{})\,\cite{Dittmaier:2003ej,Dawson:2003kb}, the latter at
next-to-\nlo{} (\nnlo{}) \qcd{}\,\cite{Harlander:2003ai}---enter with
Higgs-mass dependent weights. For larger Higgs mass, the logarithms
discussed above become more important, so the \fs{5} is expected to
provide the more reliable result, and thus receives a larger
weight. This is indeed confirmed by approaches aiming at a theoretically
better-founded matching of the underlying
processes\,\cite{Maltoni:2012pa,Forte:2015hba,Bonvini:2015pxa}.\footnote{For
  a comparison of differential distributions in $\bbh$ production based
  on the \fs{4} and the \fs{5}, see \citere{Wiesemann:2014ioa}.}

Due to the small value of the charm-quark mass $\mcharm\sim 1$\,GeV, a
charm-initiated approach for the calculation of the total inclusive
cross section, $c\bar c\to \phi$, is preferable over a 3-flavor scheme
(\fs{3}) description (\lo{} process $gg\to c\bar c\phi$) already for
much smaller values of the Higgs boson mass.  It can be evaluated both
in the \fs{4} and the \fs{5}, where in the latter case the bottom quark
plays the role of a spectator. Since, as we will show below,
interference effects involving the bottom and the charm Yukawa coupling
are absent, the only technical difference in evaluating the \fs{4} and
the \fs{5} result for $\sigma(c\bar c\to \phi)$ is a change of the
\pdf{} set. All results in this paper are obtained in the \fs{5}.

Analogous considerations apply to other quark-associated production
modes.  As we will show in this paper, the corresponding \nnlo{} {\it
  partonic} cross sections differ only by an obvious overall factor,
given by the ratio of the respective Yukawa couplings, as long as the
dynamical quark masses (as opposed to the Yukawa couplings) are
neglected. The latter condition is anyway necessary in a partonic
formulation of these scattering processes. 

We can therefore use the known partonic \nnlo{} results for the process
$b\bar b\to \phi$\,\cite{Harlander:2003ai}, and translate them into
hadronic cross sections for arbitrary initial-state quarks.  This will
be explained in more detail in the next section.
Section\,\ref{sec:numerics} uses these results to determine the central
renormalization and factorization scales for all heavy-quark initiated
Higgs production processes, and provides theoretical predictions through
\nnlo{}. In addition, the impact of hard radiation is disentangled from
the purely \pdf{}-induced effects. Section\,\ref{sec:conclusions}
contains our conclusions.

\section{Calculation}\label{sec:calc}

\begin{figure}
  \begin{center}
    \begin{tabular}{c}
      \raisebox{0em}{\includegraphics[viewport=150 600 300
          700,height=.13\textheight,clip]{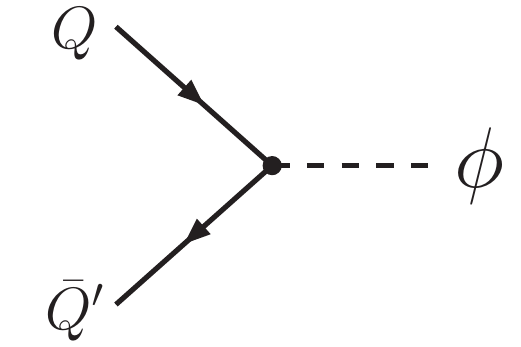}}
    \end{tabular}
    \parbox{.9\textwidth}{
      \caption[]{\label{fig:qqhlo}\sloppy \lo{} Feynman diagram for
        $\Q\bar \Q'\to \phi$, defining the process $pp\to \qqh$ in the \fs{5}.
    }}
  \end{center}
\end{figure}

We denote by $\qqh$ the process for the associated production of a Higgs
boson $\phi$ with a $\Q'\bar \Q$ pair in the \fs{5}, whose \lo{} Feynman
diagram is given by \fig{fig:qqhlo}. Depending on the specific flavors
of $\Q$ and $\Q'$, $\phi$ can be electrically neutral or charged. Within
\qcd{}, renormalization of the $\qqh$ coupling, and thus also its
anomalous dimension, is independent of the quark flavors $\Q$ and $\Q'$.
Since we work in the massless-quark limit throughout this paper, the
underlying theory is chirally symmetric, which means that all our
results apply to scalar as well as pseudo-scalar Higgs bosons $\phi$
(see also \citere{Harlander:2003ai}); scalar/pseudo-scalar interference
terms vanish.

At \nlo{} \qcd{}, aside from the virtual corrections to the \lo{}
process, the real radiation processes $\Q\bar\Q'\to g\phi$, $g\Q\to
\Q'\phi$, and $g\bar \Q'\to \bar \Q\phi$ need to be taken into account
in the calculation of the total cross section.  Similarly, at \nnlo{}
\qcd{}, there are the two-loop virtual corrections to the \lo{} process,
and the one-loop virtual corrections to the \nlo{} real-emissions
processes. In addition, double-real emission processes occur. Those with
two external gluons are: $\Q\bar \Q'\to gg\phi$, $\Q g\to \Q g\phi$,
$\bar\Q'g\to \bar\Q'g\phi$, $gg\to \bar \Q\Q'\phi$. The squared
amplitude composed of these processes contains only a single fermionic
trace.

\begin{figure}
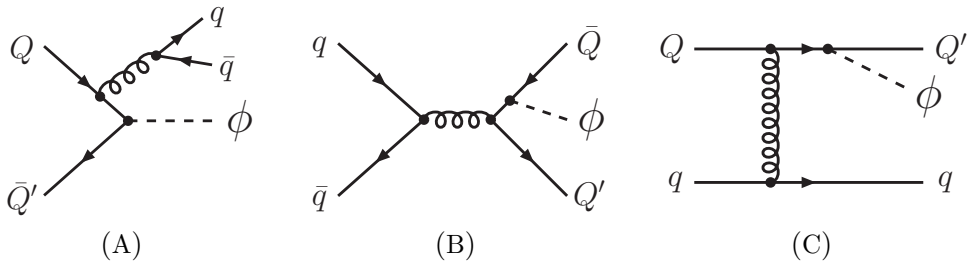

  \begin{center}
    \begin{tabular}{ccc}
      \raisebox{0em}{\includegraphics[viewport=150 470 300
          585,height=.15\textheight,clip]{pics/dias.pdf}} &
      \raisebox{0em}{\includegraphics[viewport=160 340 330
          445,height=.14\textheight,clip]{pics/dias.pdf}} &
      \raisebox{.5em}{\includegraphics[viewport=160 210 330
          300,height=.12\textheight,clip]{pics/dias.pdf}}\\
      (\myA) & (\myB) & (\myC)
    \end{tabular}
    \parbox{.9\textwidth}{
      \caption[]{\label{fig:abc}\sloppy \nnlo{} contributions to the
        $\qqh$ process which involve four external quarks. (\myC) is a
        representative for three more diagrams which are obtained by
        replacing $q\to \bar q$, or $(\Q,\Q')\to (\bar \Q',\bar \Q)$, or
        both.  }}
  \end{center}
\end{figure}

This may be different for processes with four external quarks. Their
amplitudes are given by
\begin{itemize}
\item[\myA:] $\Q\bar \Q'\to q\bar q\phi$
\item[\myB:] $q\bar q\to \Q'\bar \Q\phi$
\item[\myC:] $\Q q\to \Q' q\phi$,\quad  $\bar \Q' q\to \bar \Q q\phi$,\quad
  $\Q\bar q\to \Q'\bar q\phi$,\quad  $\bar \Q'\bar q\to \bar \Q\bar q\phi$
\end{itemize}
and correspond to the Feynman diagrams shown in \fig{fig:abc}.  Here,
$q$ denotes a quark of arbitrary flavor, and $\bar q$ the corresponding
anti-quark. The square of each of these processes involves two fermionic
traces, one of which contains both Higgs couplings.

Let us now look at potential interference terms.  If
$q\not\in\{\Q,\Q'\}$, the initial and final states of \myA, \myB, and
\myC\ are different, and they obviously cannot interfere. If $q\in
\{\Q,\Q'\}$, there are \myA\myC\ and \myB\myC\ interference terms, which
involve a single fermionic trace.

All contributions above are independent of the specific quark flavors
$\Q$ and $\Q'$.  For $\Q=\Q'=q$, however, it seems that also \myA\ and
\myB\ interfere with each other, leading to a term with two fermionic
traces, each of which contains one Higgs coupling. However, in the limit
of zero quark masses, the traces are over an odd number of Dirac
matrices and vanish.

In conclusion, aside from an overall {\it constant} Yukawa
factor,\footnote{Recall that the anomalous dimension of the $\qqh$
  vertex is independent of $\Q$ and $\Q'$, see above.} the \nnlo{}
partonic cross section for the process $\qqh$ is independent of the
quark flavors $\Q$ and $\Q'$, as long as quark masses are neglected.
Along the same lines, one observes that, for $\Q\neq\Q'$, any
interference terms between $\Q\bar \Q$- and $\Q'\bar \Q'$- initiated
Higgs production vanishes through \nnlo{}. 

Let us remark that in the analogous case of Drell-Yan production,
i.e.\ $\phi=V\in\{W,Z\}$, the \myA\myB\ interference term, which exists only
for $Z$-production, is {\it not} zero. The double-quark emission
corrections for $W$-production are therefore different from those of
$Z$-production\,\cite{Hamberg:1991np}.\footnote{This effect adds to the
  difference between $W$- and $Z$-production arising from other
  contributions, see \citere{Hamberg:1991np} for more details. Note that
  the same discussion also applies to the Higgs-Strahlung process,
  $pp\to VH$. In this case, however, there is a much more important
  difference between $V=W$ and $V=Z$ arising from the gluon induced
  $gg\to HZ$ process\,\cite{Kniehl:1990iv,Dicus:1988yh,Brein:2003wg}.}

It follows that the hadronic $\qqh$ cross section for the collision of
hadrons $h_1$ and $h_2$ can be obtained by simply convolving the \fs{5}
partonic cross section for $\bbh$ production with the appropriate
\pdf{}s.  For example, we may define
\begin{equation}
\begin{split}
\sigma_{b\bar b}(f,f') = \left[f_1\otimes f'_2 +
  f'_1\otimes f_2\right]\otimes \hat\sigma_{b\bar b}\,,
\label{eq:sigqqx1}
\end{split}
\end{equation}
where $\hat\sigma_{b\bar b}=\hat\sigma_{b\bar b}(\mhiggs^2/\hat s)$ is
the partonic cross section for the \sm{} process $b\bar b\to H+X$, which can
be found in \citere{Harlander:2003ai}, $\hat s$ is the partonic
center-of-mass energy, and $\otimes$ denotes the convolution
\begin{equation}
\begin{split}
(f\otimes g)(x) =\int_0^1\dd x_1\int_0^1\dd x_2
  f(x_1)g(x_2)\delta(x-x_1x_2)\,.
\end{split}
\end{equation}
Furthermore, $f_j(x)$ and $f'_j(x)$ are the parton densities in the
hadron $h_j$, with $f,f'\in \{q,\bar q,g\}$ and
$q\in\{d,u,s,c,b\}$. The component of the hadronic $\qqh$ cross
section which is induced by the partonic $\Q\bar \Q'$ initial state can
then be written as
\begin{equation}
\begin{split}
\sigma(\Q\bar \Q'\to \phi+X) = \betaqq\,\sigma_{b\bar b}(\Q,\bar
\Q')\,,
\end{split}
\end{equation}
where $\betaqq$ is the squared ratio of the $\Q\bar\Q'\phi$ and the
\sm{} $\bbH$ coupling. In particular, we have $\sigma(b\bar b\to H+X) =
\sigma_{b\bar b}(b,\bar b)$.

Similarly, we can define 
\begin{equation}
\begin{split}
\sigma_{bg}(f,f') &=
\left[(f_1+f'_1)\otimes g_2 +
  g_1\otimes (f_2+f'_2)\right]\otimes\hat\sigma_{bg}\,,\\
\sigma_{bb}(f,f') &= \left[f_1\otimes f_2 +
  f'_1\otimes f'_2\right]\otimes \hat\sigma_{bb}\,,\\
\sigma_{bq}(f,f') &= \bigg[ (f_1+f'_1)\otimes \Sigma_2 +
  \Sigma_1\otimes (f_2+f'_2)\bigg]\otimes \hat\sigma_{bq}\,,\\
\sigma_{gg} &= g_1\otimes g_2\otimes \hat\sigma_{gg}\,,\\
\sigma_{q\bar q} &= \sum_q( q_1\otimes \bar q_2 +
q_2\otimes \bar q_1 )\otimes \hat\sigma_{q\bar q}\,,
\end{split}
\end{equation}
where
\begin{equation}
\begin{split}
\Sigma_i = \sum_q(q_i+\bar q_i) - f_i-f'_i\,,
\end{split}
\end{equation}
and the sum runs over all quark flavors $q$. The partonic cross sections
$\hat\sigma_{ij}$ on the right hand side are $ij$-initiated components
of the partonic \sm{} $\bbH$ cross section; explicit expressions can be
found in \citere{Harlander:2003ai}.

In this way, we can calculate
\begin{equation}
\begin{split}
\sigma(\Q g\to \phi+X) + \sigma(\bar \Q' g\to \phi+X) 
&= \betaqq \sigma_{bg}(\Q,\bar \Q')\,,\\
\sigma(\Q\Q\to \phi+X) + \sigma(\bar \Q'\bar \Q'\to \phi+X) 
&= \betaqq \sigma_{bb}(\Q,\bar \Q')\,,\\
\sigma(\Q q\to \phi+X) + \sigma(\bar \Q' q\to \phi+X)
&= \betaqq \sigma_{bq}(\Q,\bar \Q')\,,\\
\sigma(gg\to \phi+X) &= \betaqq\sigma_{gg}\,,\\
\sigma(q\bar q\to \phi+X) &= \betaqq\sigma_{q\bar q}\,,
\label{eq:sigqqx2}
\end{split}
\end{equation}
where $q$ may be any (anti-)quark except $\Q$ or $\bar \Q'$.
The total inclusive hadronic cross section is then given by the sum of
all the terms in \eqns{eq:sigqqx1} and \noeqn{eq:sigqqx2}.

The implementation of this result in {\tt
  bbh@nnlo}\,\cite{Harlander:2003ai} (which is now part of
\sushi{}\,\cite{Harlander:2012pb}) is straightforward and will be
publically available in the next version of \sushi{}.\footnote{Watch
  \href{http://sushi.hepforge.org/}{\tt http://sushi.hepforge.org/}, or
  follow \href{https://twitter.com/sushi4physics}{\tt @sushi4physics on
    Twitter}.}

\section{Numerical results}\label{sec:numerics}

\subsection{Determination of the central scales}\label{sec:scales}

\begin{figure}
  \begin{center}
    \begin{tabular}{cc}
      \includegraphics[viewport=20 0 300 220,height=.25\textheight]{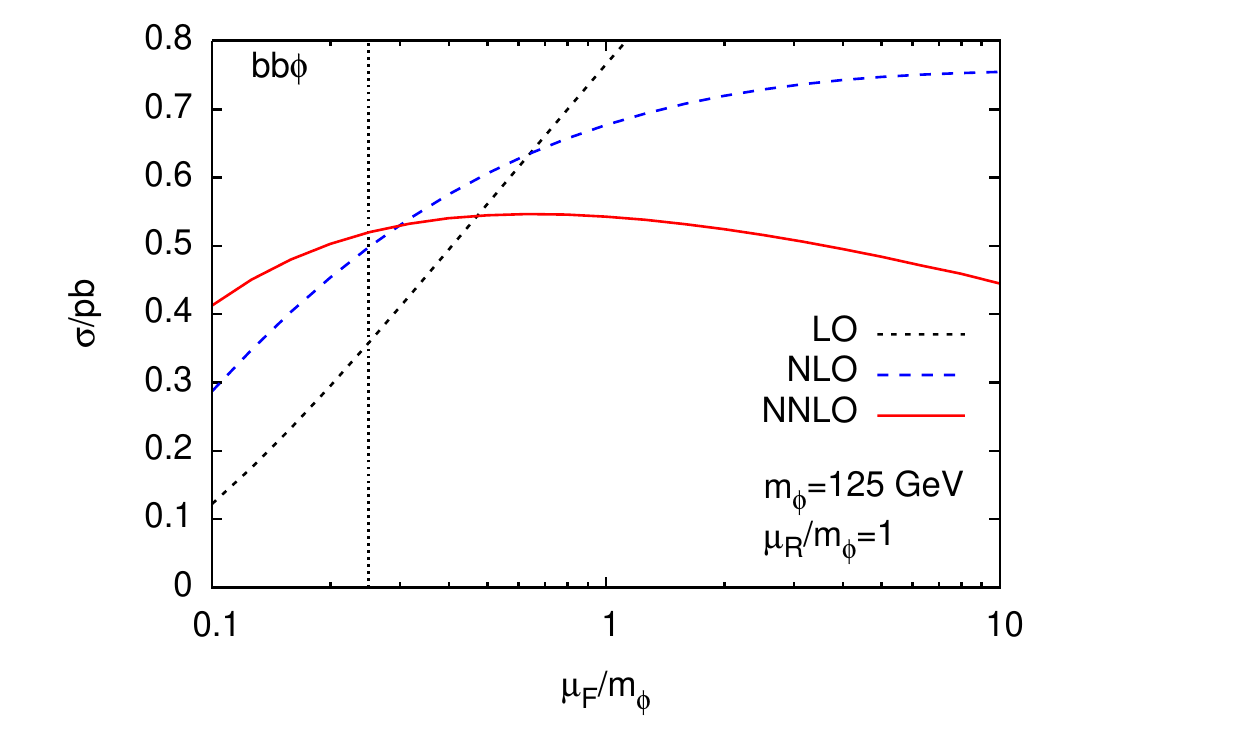}
      &
      \includegraphics[viewport=20 0 300 220,height=.25\textheight]{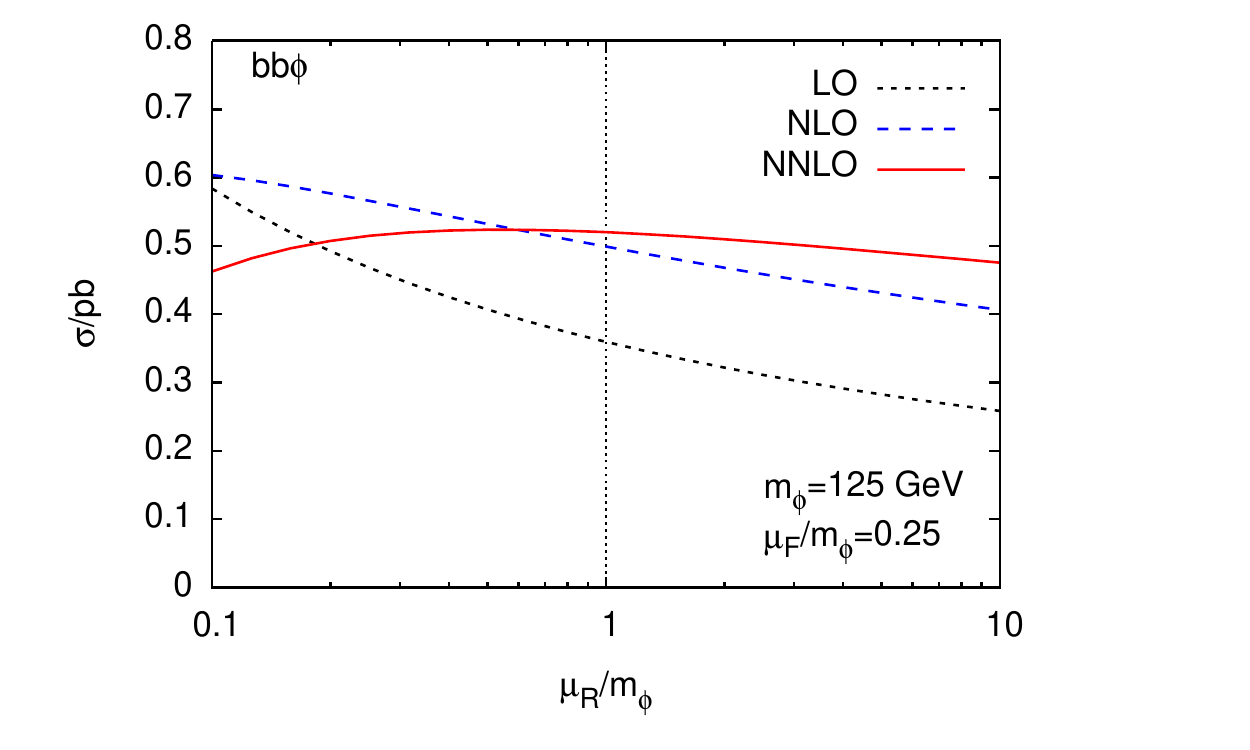}
      \\
      \includegraphics[viewport=20 0 300 220,height=.25\textheight]{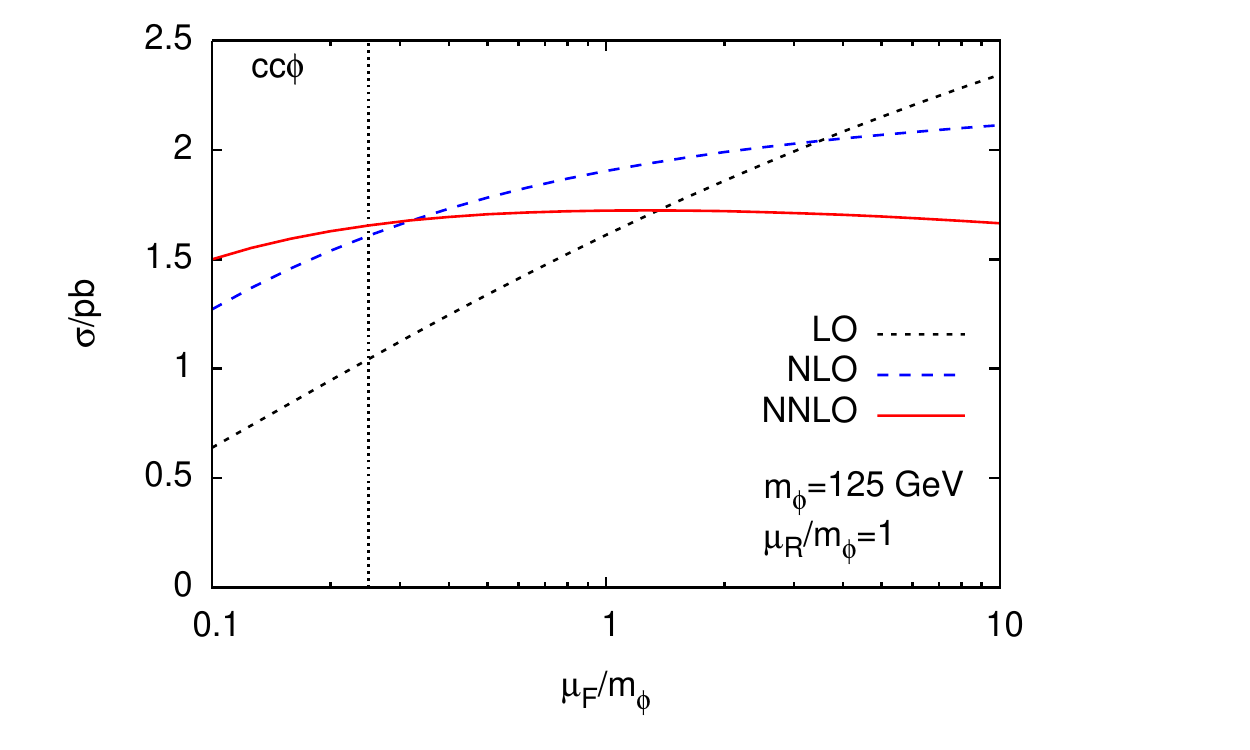}
      &
      \includegraphics[viewport=20 0 300 220,height=.25\textheight]{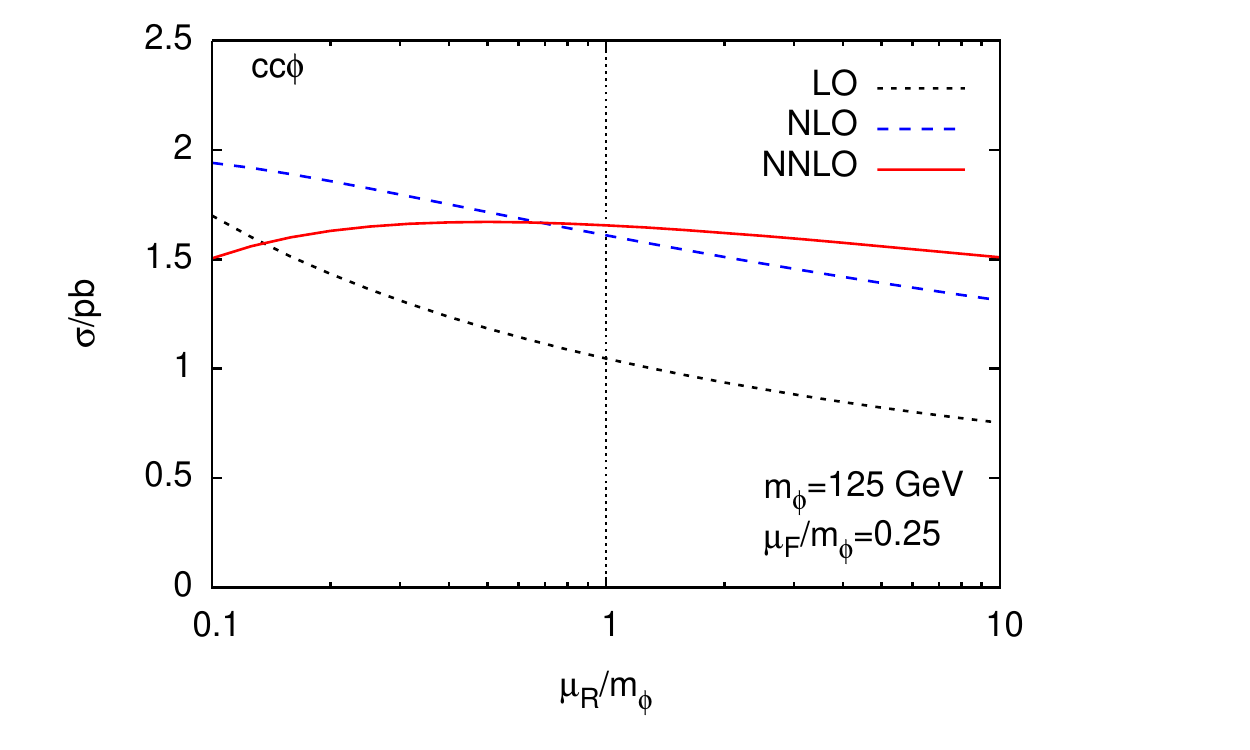}
      \\
      \includegraphics[viewport=20 0 300 220,height=.25\textheight]{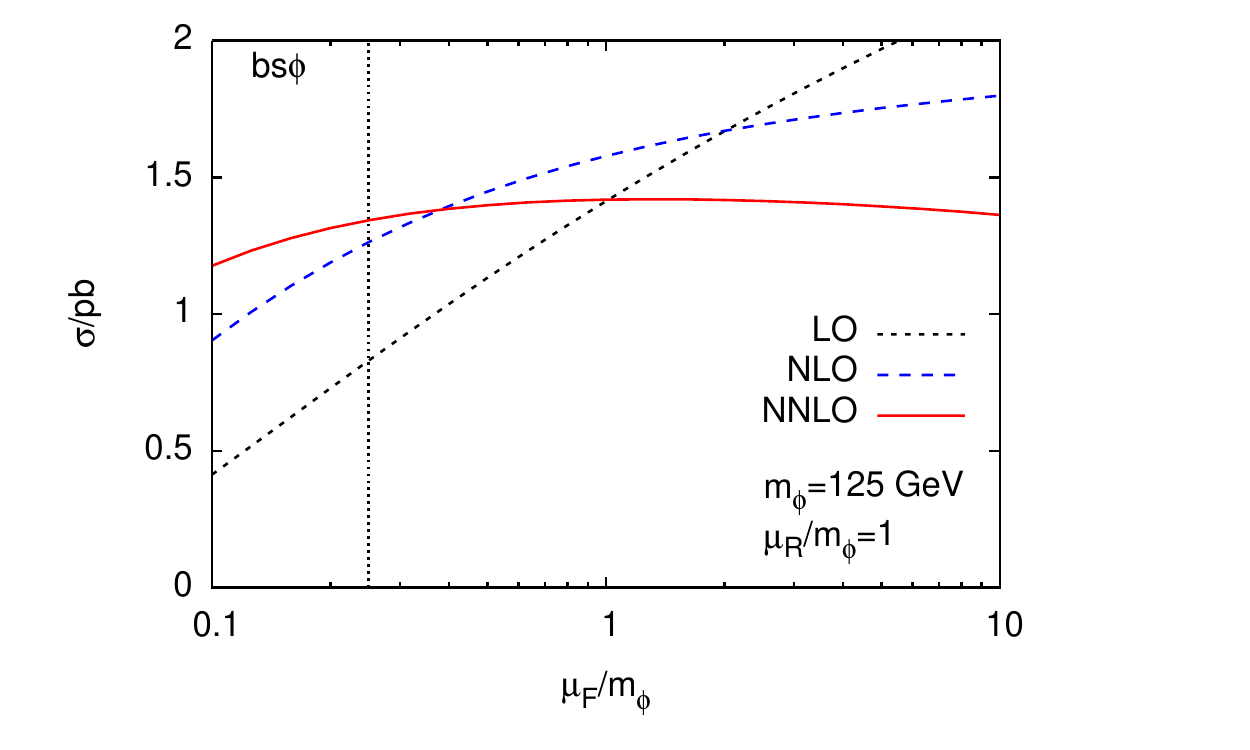}
      &
      \includegraphics[viewport=20 0 300 220,height=.25\textheight]{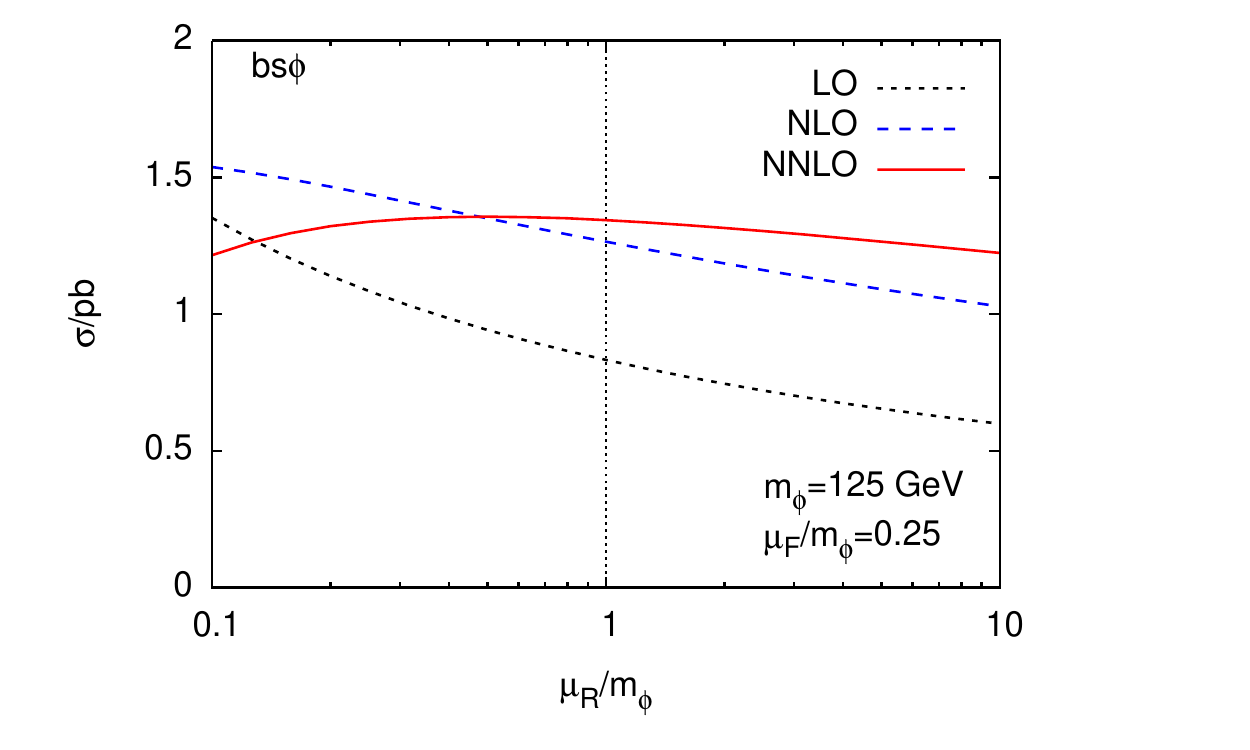}
      \\
    \end{tabular}
    \parbox{.9\textwidth}{
      \caption[]{\label{fig:qqh125}\sloppy \lo{} (black dots), \nlo{}
        (blue dashes), and \nnlo{} result (solid red) for the total
        cross sections of the processes $\bbh$, $\cch$, and $\bsh$ (top
        to bottom) at $\mphi=125$\,GeV. Left column: $\muF$-dependence
        for $\muR=\mphi$: right column: $\muR$-dependence for
        $\muF=\mphi/4$. The vertical dotted lines at $\muFhat=1/4$
        (left) and $\muRhat=1$ (right) are introduced to guide the
        eye. At \nklo{n} order, the corresponding central {\abbrev
          MSTW}2008 set and its associated value of $\alpha_s(M_Z)$ has
        been used; $\alpha_s(M_Z)$ and $\mbottom(\mbottom)=4.18$\,GeV
        have been evolved to $\muR$ at $(n+1)$-loop order.}}
  \end{center}
\end{figure}

As a reference, the upper two plots of \fig{fig:qqh125} show the first
three perturbative orders for the $\bbh$ cross section for
$\mphi=125$\,GeV and $\beta_{bb}=1$ as a function of the factorization
scale $\muF$ (left), and the renormalization scale $\muR$ (right). These
results are well-known\,\cite{Harlander:2003ai}; they corroborate the
choice $(\muRhat,\muFhat)=(1,1/4)$ as the central values for the
scales\,\cite{Boos:2003yi,Maltoni:2003pn,Rainwater:2002hm}, where we
have introduced the normalized scales
\begin{equation}
\begin{split}
\muRhat\equiv \muR/\mphi\,,\qquad
\muFhat\equiv \muF/\mphi\,.
\end{split}
\end{equation}

We may formalize the justification of this choice by considering the
variation $\Delta_\text{F}$ of the \nnlo{} hadronic cross section
$\sigma$ within the interval $\muRhat\in [1/10,10]$, while fixing
$\muFhat$:
\begin{equation}
\begin{split}
\Delta_\text{F} =
2\,\frac{\max\sigma-\min\sigma}{\max\sigma+\min\sigma}\bigg|_{\muFhat}\,.
\end{split}
\end{equation}
The central factorization scale $\muFhat^{(0)}$ can then be defined as
the value of $\muFhat$ where $\Delta_\text{F}$ is minimal. The analogous
procedure (with {\abbrev R}$\leftrightarrow${\abbrev F}) can be used to
define the central renormalization scale $\muRhat^{(0)}$.

We performed this study for all heavy-quark initiated processes by
calculating $\sigma$ on an equidistant $21\times21$ logarithmic grid in
the $(\muRhat,\muFhat)$ plane, i.e., using the values
$\muRhat,\muFhat\in\{10^{n/10}\,,\ n=-10,-9,\ldots,9,10\}$. When quoting
numbers, we will round these values to two significant digits; e.g., we
will refer to $\muFhat=10^{-3/5}=0.2512\ldots$ simply as
$\muFhat=0.25$, or $\muFhat=1/4$ for that matter.

For $\mphi=125$\,GeV, we find $\muFhat^{(0)}=1/4$ in this way,
independent of the quark flavors $\Q$ and $\Q'$. This is an interesting
observation, because this value has been derived specifically for
$\Q=\Q'=b$ using kinematical
considerations\,\cite{Boos:2003yi,Maltoni:2003pn,Rainwater:2002hm};
  the fact that all other quark-initiated processes seem to favor the
  same $\muFhat^{(0)}$ is not at all obvious from these discussions.
  Following the above procedure, the central renormalization scale turns
  out to be $\muRhat^{(0)} = 0.79$ for the $\bbh$ process, while for the
  other quark flavors we find $\muRhat^{(0)} = 0.63$.

For $\mphi=600$\,GeV, all processes favor an even smaller value of the
factorization scale, namely $\muFhat^{(0)} = 0.16$. Also the central
renormalization scale comes out smaller: we find $\muRhat^{(0)} =
0.63$ for $\bbh$, $\muRhat^{(0)}=0.5$ for $\cch$ and $\bch$, and
$\muRhat^{(0)}=0.4$ for $\bsh$ and $\csh$.

\begin{figure}
  \begin{center}
    \begin{tabular}{cc}
      \includegraphics[viewport=20 0 300 220,height=.25\textheight]{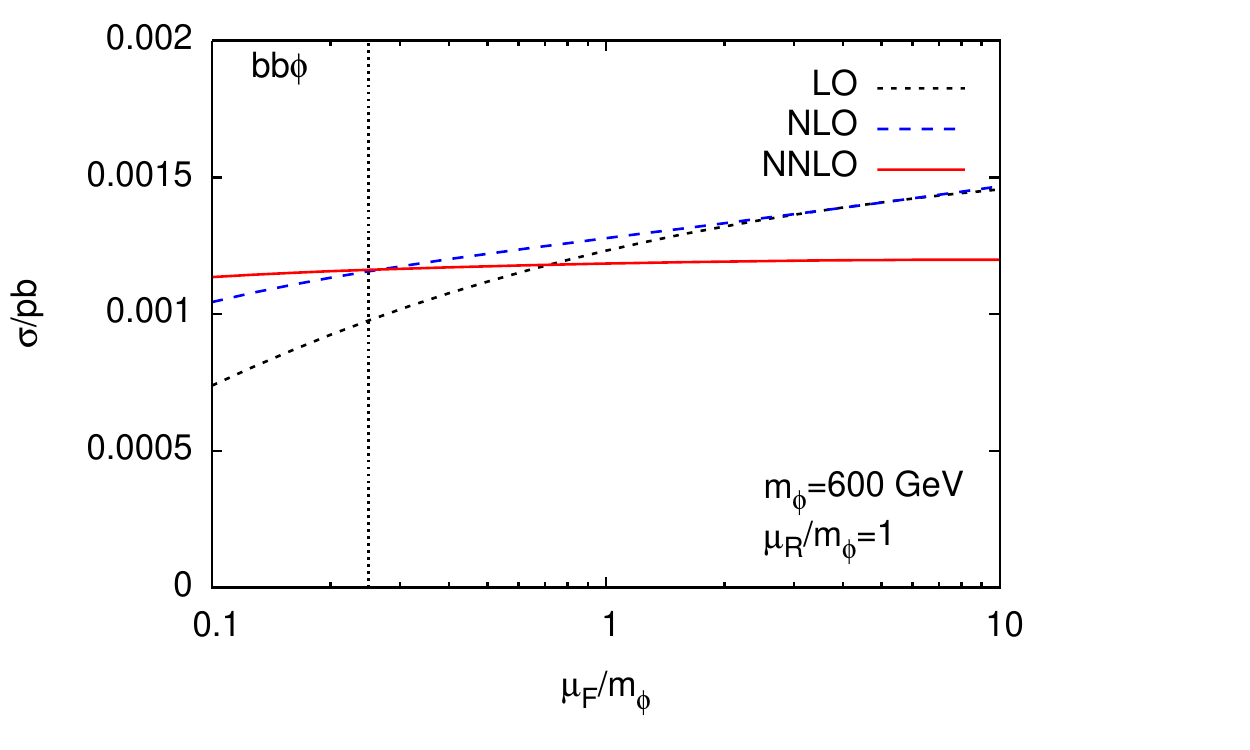}
      &
      \includegraphics[viewport=20 0 300 220,height=.25\textheight]{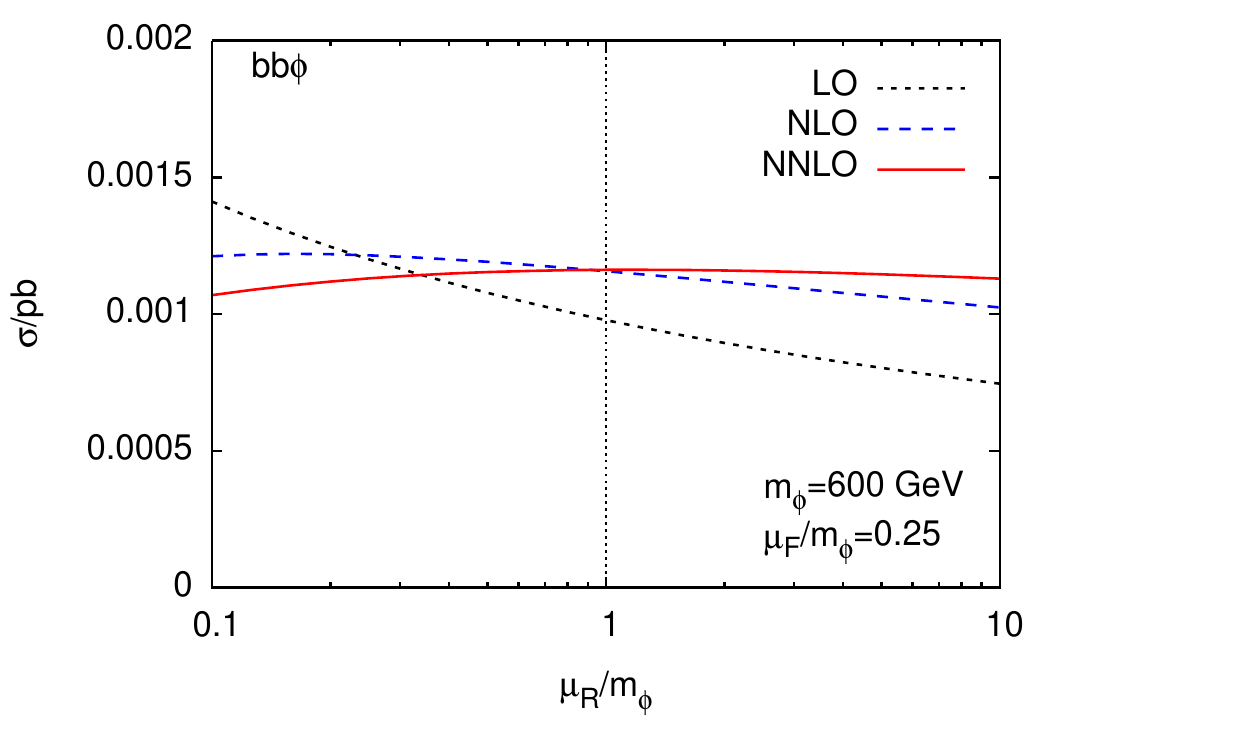}
      \\
      \includegraphics[viewport=20 0 300 220,height=.25\textheight]{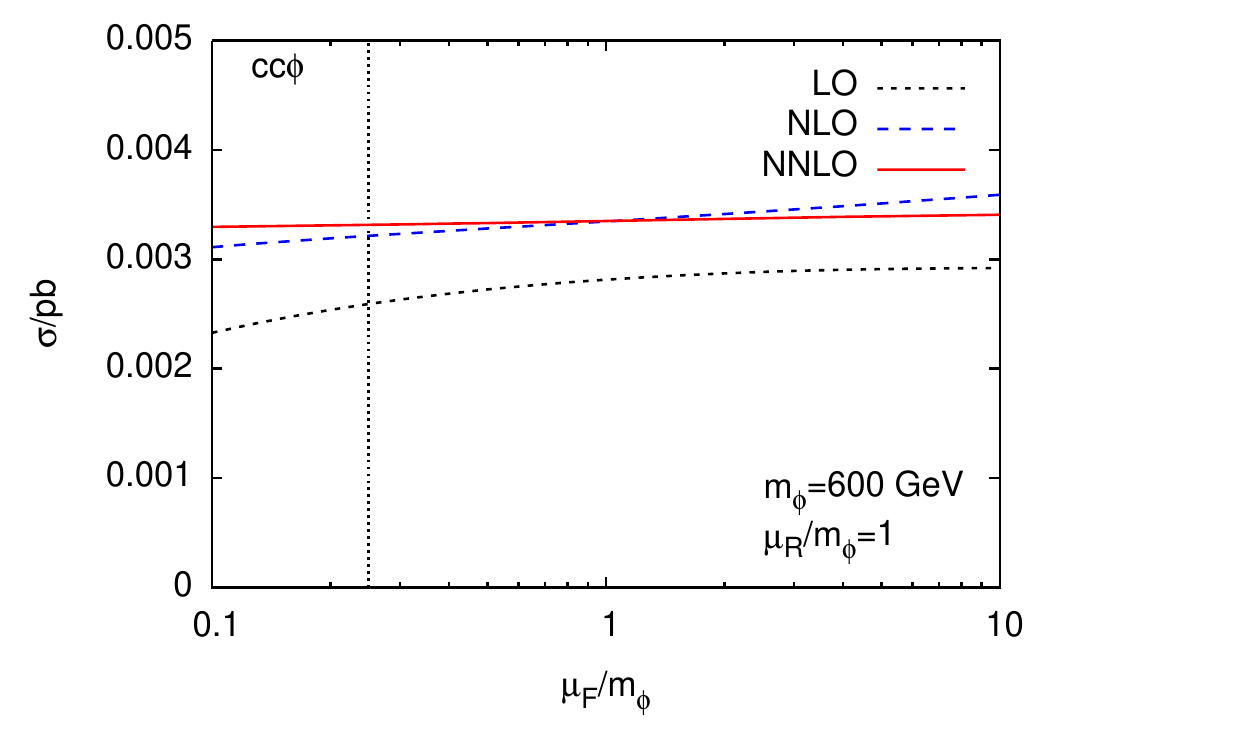}
      &
      \includegraphics[viewport=20 0 300 220,height=.25\textheight]{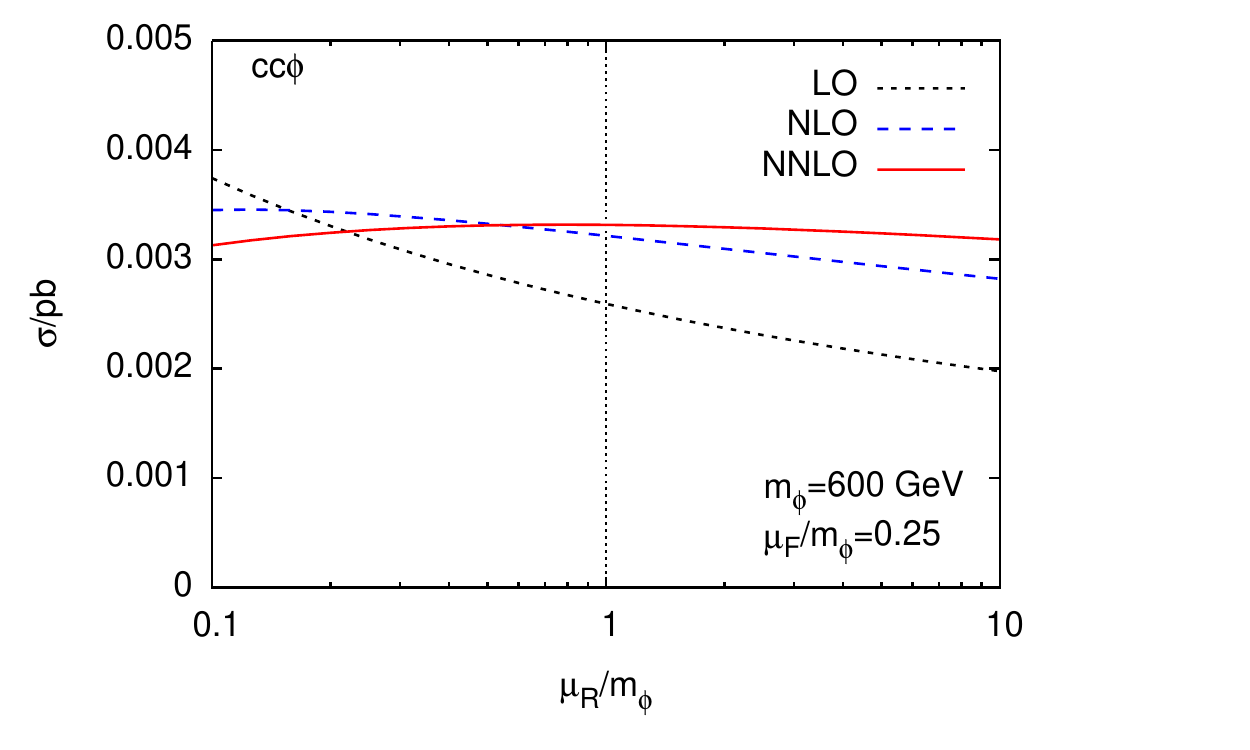}
      \\
      \includegraphics[viewport=20 0 300 220,height=.25\textheight]{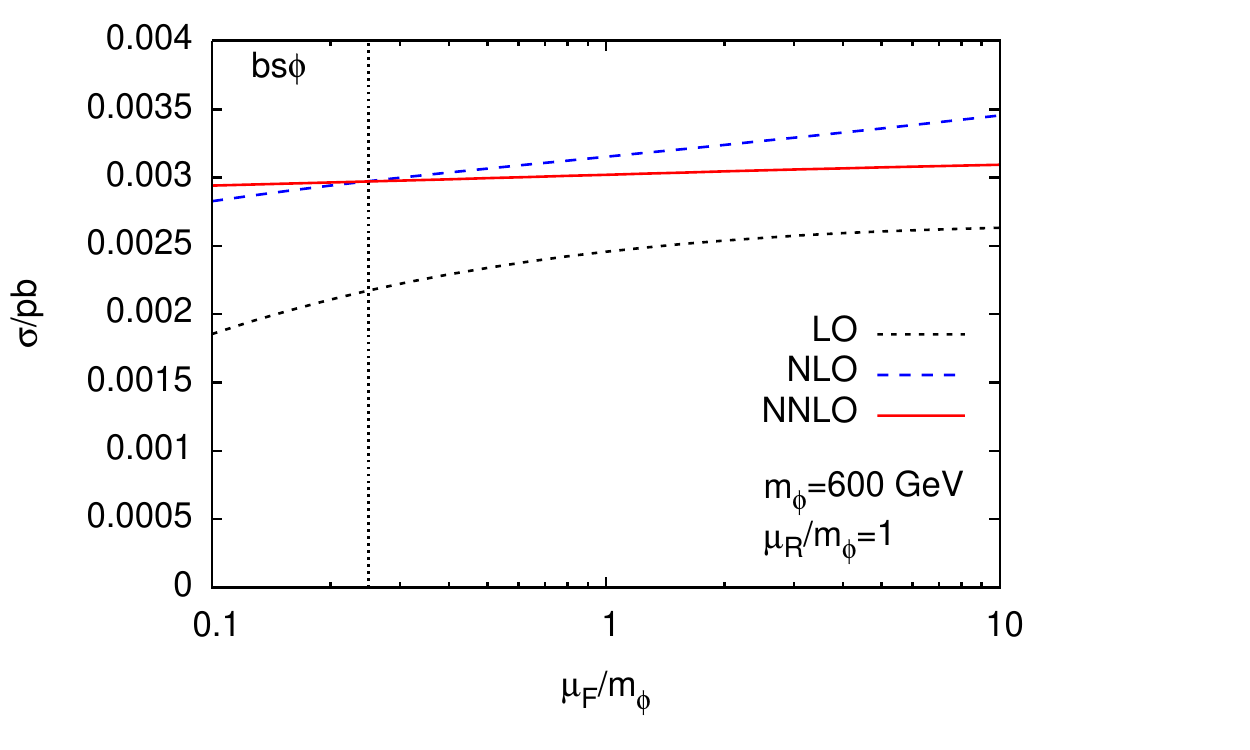}
      &
      \includegraphics[viewport=20 0 300 220,height=.25\textheight]{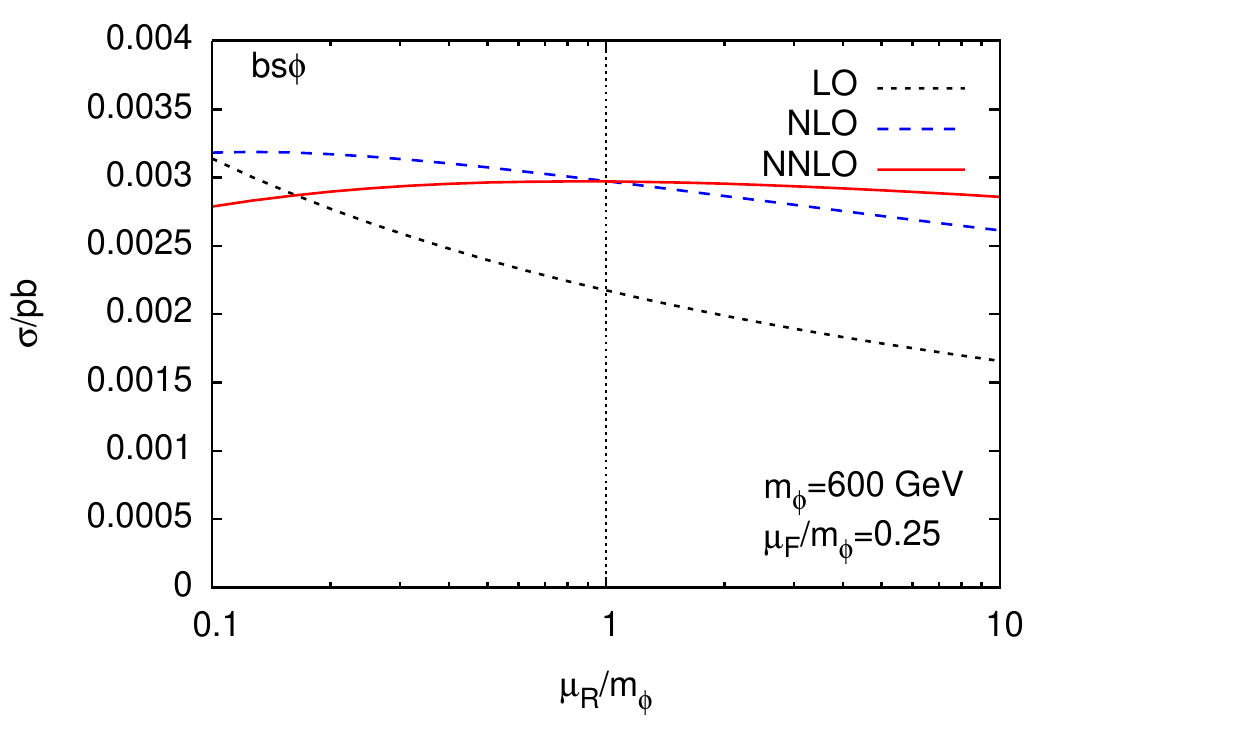}
      \\
    \end{tabular}
    \parbox{.9\textwidth}{
      \caption[]{\label{fig:qqh600}\sloppy
        Same as \fig{fig:qqh125}, but for $\mphi=600$\,GeV.
        }}
  \end{center}
\end{figure}

However, in all cases, the minima are sufficiently shallow to justify
also the choice $(\muRhat^{(0)},$ $\muFhat^{(0)}) = (1,1/4)$.  Exemplary
plots for the $\bbh$, $\cch$, and $\bsh$ processes are shown in
\fig{fig:qqh125} for $\mphi=125$\,GeV, and in \fig{fig:qqh600} for
$\mphi=600$\,GeV. All cross sections correspond to $\betaqq=1$, i.e.,
the Yukawa coupling is assumed identical to the one for \sm{} $\bbH$
production. For the \nklo{n} curve, it is evaluated from
$\mbottom(\mbottom)=4.18$\,GeV by $(n+1)$-loop evolution with $n_f=5$
active flavors to\footnote{The notation $m_q^{(n_f)}$ indicates that
  $m_q$ is renormalized in $n_f$-flavor \qcd{}.} $\mbottom(\muR)\equiv
\mbottom^{(5)}(\muR)$. Thus, in order to derive the $\cch$ cross section
within the \sm{}, for example, the plots in the second rows of
\figs{fig:qqh125} and \ref{fig:qqh600} should be scaled by
$\beta_{cc}=(\mcharm^{(5)}(\mbottom)/\mbottom^{(5)}(\mbottom))^2\approx
0.049$, where we have used 4-loop running to determine
$\mcharm^{(5)}(\mbottom)=0.926$\,GeV from
$\mcharm^{(4)}(3\,\text{GeV})=0.986$\,GeV\,\cite{Chetyrkin:2000yt}. In
the \sm{}, the $\cch$ cross section is therefore about 6-7 times smaller
than the $\bbh$ cross section.  All plots have been produced with the
{\tt MSTW2008} \pdf{} sets\,\cite{Martin:2009iq} as implemented in the
{\tt LHAPDF} library\,\cite{Buckley:2014ana,lhapdf}, and the associated
value of $\alpha_s(M_Z)=0.139/0.120/0.117$ at
\lo{}/\nlo{}/\nnlo.

Recall that the role of the central values is to determine the position
of a ``reasonable'' interval for $\muFhat$ and $\muRhat$; the variation
of the cross section within this interval should then give a clue of the
associated theoretical error induced by missing higher-order effects.
Due to the unphysical nature of the renormalization and factorization
scale, any procedure to ``determine'' their central values is formally
arbitrary, though not necessarily sensible. The fact that, at
$(\muRhat,\muFhat)=(\muRhat^{(0)},\muFhat^{(0)})$, the \nnlo{}
corrections are significantly smaller than the \nlo{} ones in all cases
studied here (see \figs{fig:qqh125} and \ref{fig:qqh600}), confirms that
the procedure defined above is indeed sensible. Other observations
concerning the choice of the central scale in the case of the $\qqh$
processes will be recalled in Section\,\ref{sec:philo}.

\clearpage

\subsection{Cross sections and uncertainties}\label{sec:xsecunc}

Numerical values for the cross sections at the \lhc{} with 13\,TeV are
shown in the form $\langle\sigma\rangle\pm \Delta_\mu \pm
\Delta_\text{\pdf}$ in Table\,\ref{tab:xsecs}, where $\Delta_\mu$ and
$\Delta_\text{\pdf{}}$ are the scale and \pdf{} uncertainty,
respectively, to be defined below. Again, for all processes, we assumed
the Yukawa coupling to be equal to the \sm{} $\bbh$ coupling.  For these
numbers, we evolved $\mbottom(\mbottom)=4.18$\,GeV to $\mbottom(\mphi)$
at 4-loop level, and subsequently $\mbottom(\mphi)$ to $\mbottom(\muR)$
at 3-loop level. The numerical difference to the single-step 4-loop
evolution from $\mbottom(\mbottom)$ to $\mbottom(\muR)$ as used in all
the plots of this paper is negligible.

The {\tt MMHT2014} \pdf{} set is employed for the convolution of the
partonic cross section.\footnote{Since this section aims at presenting
  the most up-to-date cross section predictions, we make use of
  latest-generation \pdf{} sets here. The reason for using older sets in
  Section\,\ref{sec:scales} is a technical one; it has no effect on the
  values for the central scales obtained there.} At \nnlo{}, it uses
$\alpha_s(M_Z)=0.118$, and an on-shell bottom-quark mass of
$m_{b,\text{\pdf}}=\mbottom^\text{OS}=4.75$\,GeV, which is very close to
$4.78$\,GeV, the value obtained by two-loop conversion from
$\mbottom(\mbottom)=4.18$\,GeV to the o-shell scheme (one-loop
conversion gives 4.56\,GeV, three-, and four-loop conversion both lead
to 4.93\,GeV). The dependence of the cross sections on the heavy quark
masses will be discussed in more detail below.

We evaluated the cross section for the seven pairs of scales\footnote{In
  Table\,\ref{tab:xsecs}, we do {\it not} use the grid values, but
  actually set $1/4=0.25000\ldots$, etc.}
\begin{equation}
\begin{split}
(\muRhat,\muFhat) =
\{(1/2,1/8),(1/2,1/4),(1,1/8),(1,1/4),(1,1/2),(2,1/4),(2,1/2)\}\,,
\end{split}
\end{equation}
using the central \nnlo{} \pdf{} set of {\tt
  MMHT2014}\,\cite{Harland-Lang:2014zoa}.  The corresponding
maximal/minimal values of the cross section,
$\sigma_\mu^\text{max/min}$, then determine the scale error interval as
$\Delta_\mu=(\sigma_\mu^\text{max}-\sigma_\mu^\text{min})/2$ quoted in
Table\,\ref{tab:xsecs}.  The \pdf{} uncertainty $\Delta_\text{\pdf}$ is
determined from the 25 eigenvector sets of {\tt MMHT2014} while setting
$(\muRhat,\muFhat)=(1,1/4)$; the central value $\langle\sigma\rangle$ is
the mid-point of the \pdf{} error interval.

\renewcommand{\arraystretch}{1.2}
\begin{table}
\begin{center}
\begin{tabular}{c}
\input{xsecs-MMHT2014.tex}
\end{tabular}
\caption[]{\label{tab:xsecs} Cross sections for the processes $\qqh$ at
  13\,TeV. The $\qqh$ coupling is assumed equal to the \sm{}
  $\bbH$ coupling in all cases. The cross sections hold for both scalar
  and pseudo-scalar Higgs bosons. The first uncertainty is due to scale
  variation, the second one denotes the \pdf{} error (see main text for
  more details).}
\end{center}
\end{table}

\begin{figure}
  \begin{center}
    \begin{tabular}{cc}
      \includegraphics[viewport=20 0 300
        220,height=.3\textheight]{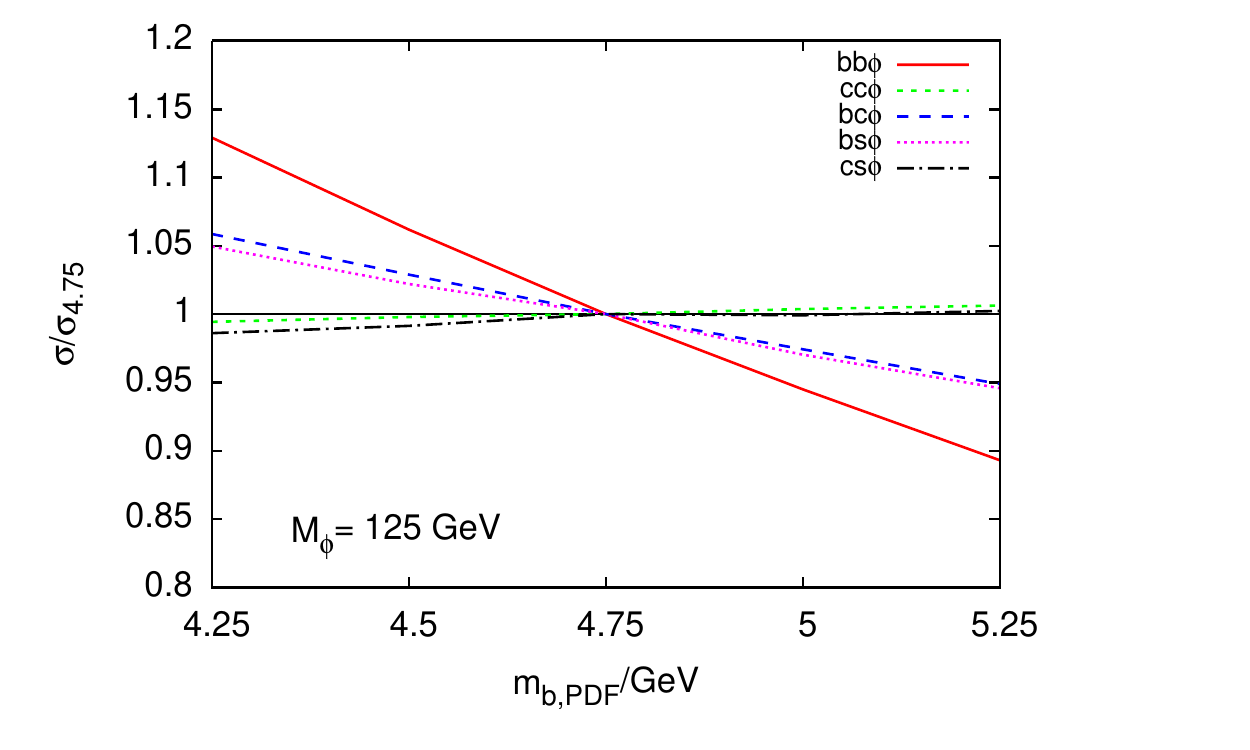} &
      \includegraphics[viewport=20 0 300
        220,height=.3\textheight]{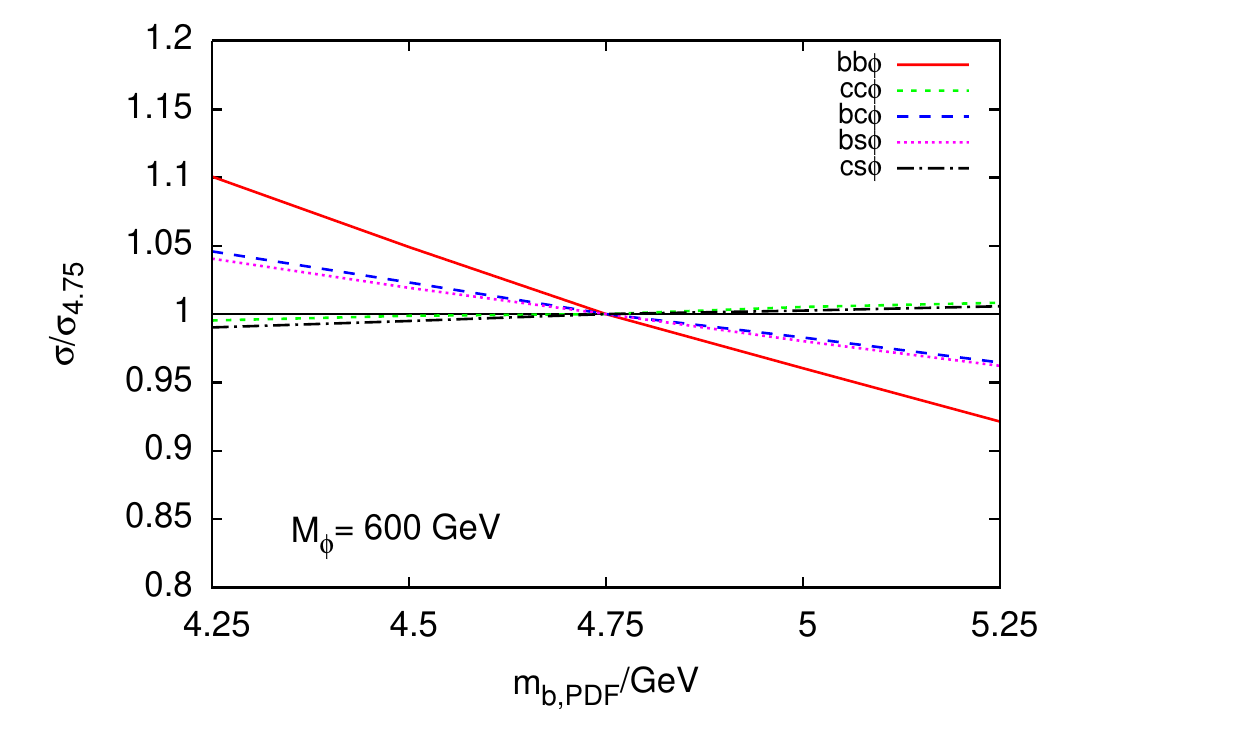}\\
      \includegraphics[viewport=20 0 300
        220,height=.3\textheight]{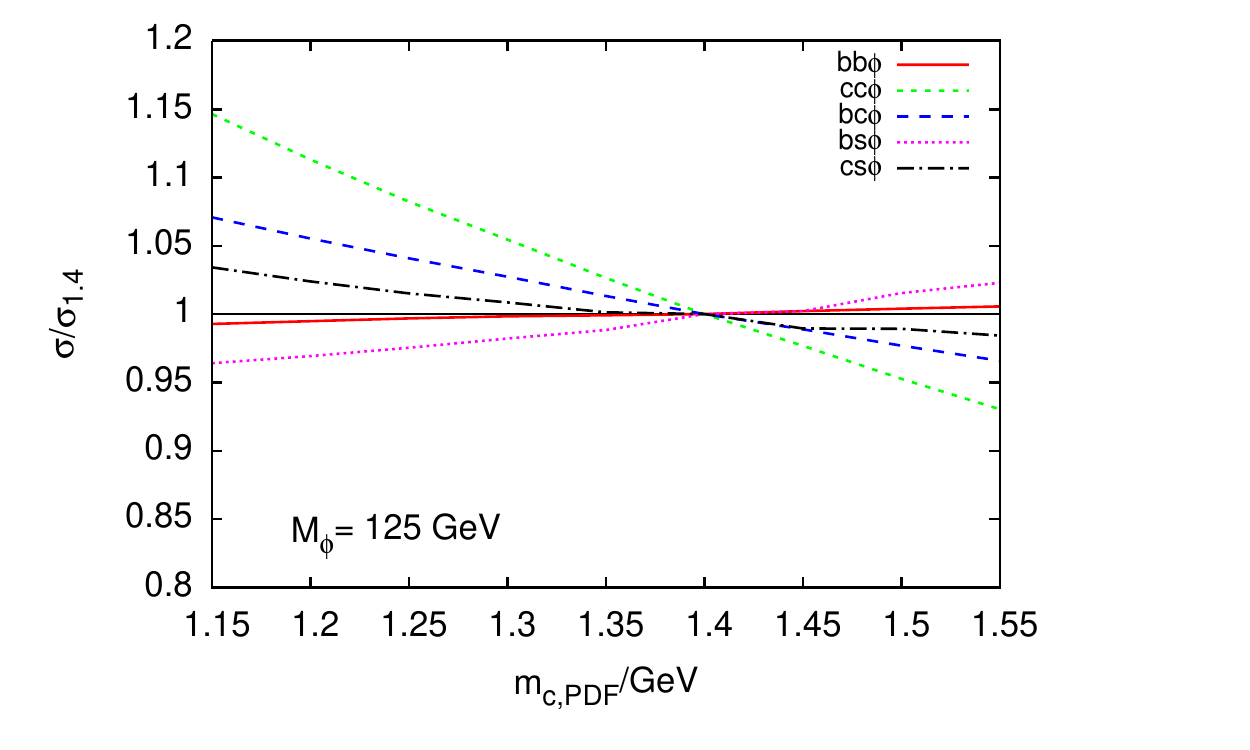} &
      \includegraphics[viewport=20 0 300
        220,height=.3\textheight]{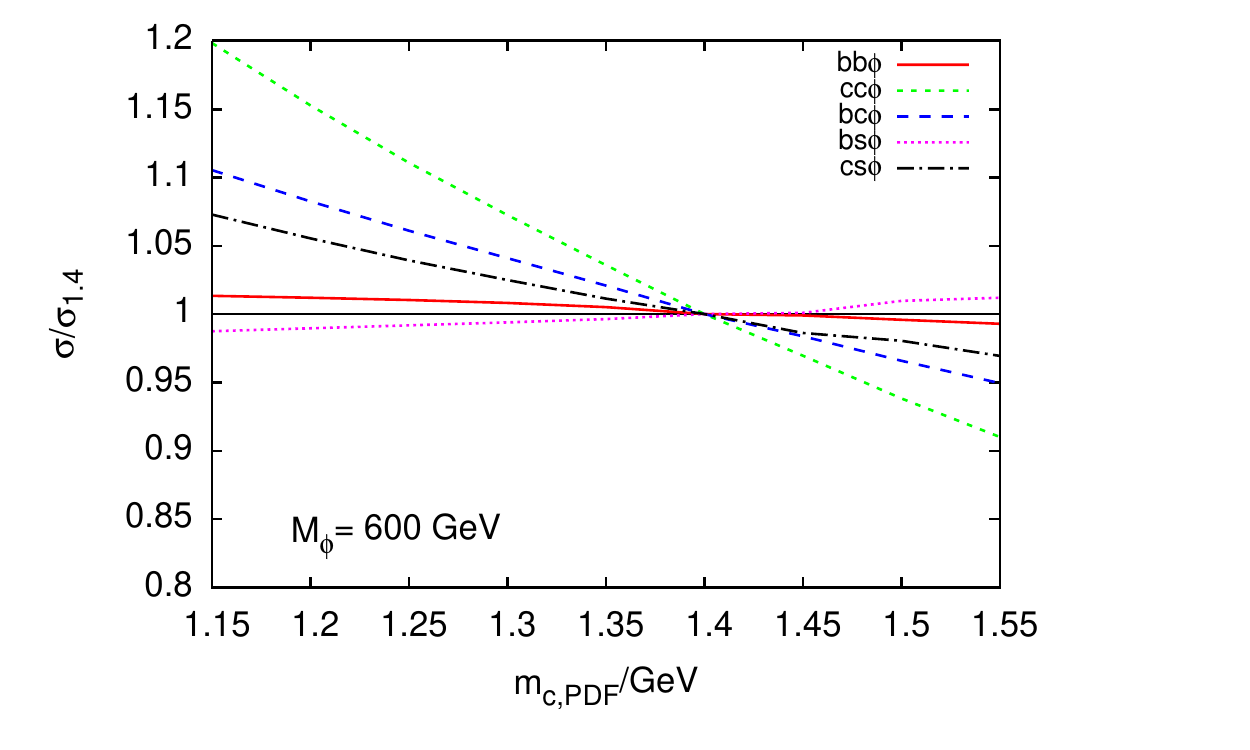}
    \end{tabular}
    \parbox{.9\textwidth}{
      \caption[]{\label{fig:mqdep}\sloppy Dependence of the hadronic
        cross section on the heavy-quark mass value $m_{q,\text{PDF}}$
        used in the \pdf{} fit (curves for {\tt MMHT2014}). Upper row:
        $q=b$; lower row: $q=c$. Left: $\mphi=125$\,GeV; right:
        $\mphi=600$\,GeV. Normalization is to the default \pdf{} set
        with $(m_{b,\text{PDF}},m_{c,\text{PDF}})=(4.75,1.4)$\,GeV. }}
  \end{center}
\end{figure}

We checked that we obtain comparable results when using the {\tt CT14}
\pdf{} set\,\cite{Dulat:2015mca}, while the default {\tt NNPDF3.0}
set\,\cite{Ball:2014uwa} typically leads to larger $\bbh$ and $\bch$
cross sections ($15$\% and $11\%$ at $\mphi=125$\,GeV, respectively),
which is most likely due to the significantly smaller bottom-quark mass
assumed in that set ($\mbottom|_\text{\tt
  NNPDF3.0}=4.18$\,GeV).  This motivates a study of the sensitivity of
the results to the heavy-quark masses $m_{b,\text{\pdf}}$,
$m_{c,\text{\pdf}}$ used in the \pdf{} fits. To that aim, we use the
{\tt mbrange\_nf5} and {\tt mcrange\_nf5} versions of {\tt MMHT2014},
whose member sets correspond to different values of the bottom and charm
quark masses.  \fig{fig:mqdep} shows the $\qqh$ cross sections
(interpolated between the disrete quark-mass values) for each of these
sets. All curves are evaluated at $(\muRhat,\muFhat)=(1,1/4)$ for fixed
Yukawa coupling (determined from $\mbottom(\mbottom)=4.18$\,GeV as
described above), and are normalized to the one with the default \pdf{}
quark mass values
$(m_{b,\text{\pdf}},m_{c,\text{\pdf}})=(4.75,1.4)$\,GeV. We observe that
the cross sections and the \pdf{} quark masses are anti-correlated, and
that a change of $m_{b,\text{\pdf}}$ by $1$\% changes the $\bbh$ cross
section by about 1.5\% at $\mphi=125$\,GeV, and a little less at
$\mphi=600$\,GeV; the $\bch$ and $\bsh$ cross sections change by about
$0.7$\%. On the other hand, the $\cch$ cross section for
$\mphi=125$\,GeV changes by about $0.9$\% when varying
$m_{c,\text{\pdf}}$ by 1\%, and a little more at $\mphi=600$\,GeV. Note
that such a variation is largely compensated by a corresponding change
of the quark mass in the Yukawa coupling (see
\citere{Bagnaschi:2014zla}).

Note that, for the $\bbh$ process, a much more thorough account of the
quark-mass effects has been obtained by a consistent matching between
the various energy regimes involved in this
process \cite{Bonvini:2015pxa,Forte:2015hba}. A similar analysis
could be performed for the general $\qqh$ processes discussed here;
however, we expect the corresponding effects to be much smaller than
other theoretical and expected experimental uncertainties for these processes.

\clearpage
\subsection{Effects due to the hard process}\label{sec:hard}

At \lo{}, the partonic $\qqh$ cross section is
\begin{equation}
\begin{split}
\hat\sigma(x) \sim \delta(1-x)\,,\quad x=\mphi^2/\hat s\,,
\end{split}
\end{equation}
and thus the \lo{} hadronic
cross section for $\qqh$ production is proportional to the \lo{} $\bbh$
cross section,
\begin{equation}
\begin{split}
\sigma^\text{\lo}(\qqh) = \frac{\betaqq{\cal E}(\Q',\bar \Q)}{{\cal
    E}(b,\bar b)} \sigma^\text{\lo}(\bbh)\,,
\label{eq:lumirat}
\end{split}
\end{equation}
where 
\begin{equation}
\begin{split}
  {\cal E}(f,f') &\equiv f_1\otimes f'_2+ f'_1\otimes f_2
\end{split}
\end{equation}
is the $ff'$ parton luminosity in $h_1h_2$ collisions. To a first
approximation, one might be tempted to apply the rescaling of
\eqn{eq:lumirat} also at higher orders. In order to see to what extent
such an approximation is valid, let us study the double ratio
\begin{equation}
\begin{split}
R_{\qqh} &= \frac{\sigma(\qqh)/\sigma(\bbh)}{\betaqq{\cal E}(\Q',\bar
  \Q)/{\cal E}(b,\bar b)}
\label{eq:rqqh}
\end{split}
\end{equation}
at \nlo{} and \nnlo{} \qcd{}, where at each order it is understood to
use the appropriate set of \pdf{}s, e.g., \pdfnlo\ at \nlo{}, and
\pdfnnlo\ at \nnlo{}.

\begin{figure}
  \begin{center}
    \begin{tabular}{cc}
      \includegraphics[viewport=20 0 300 220,height=.25\textheight]{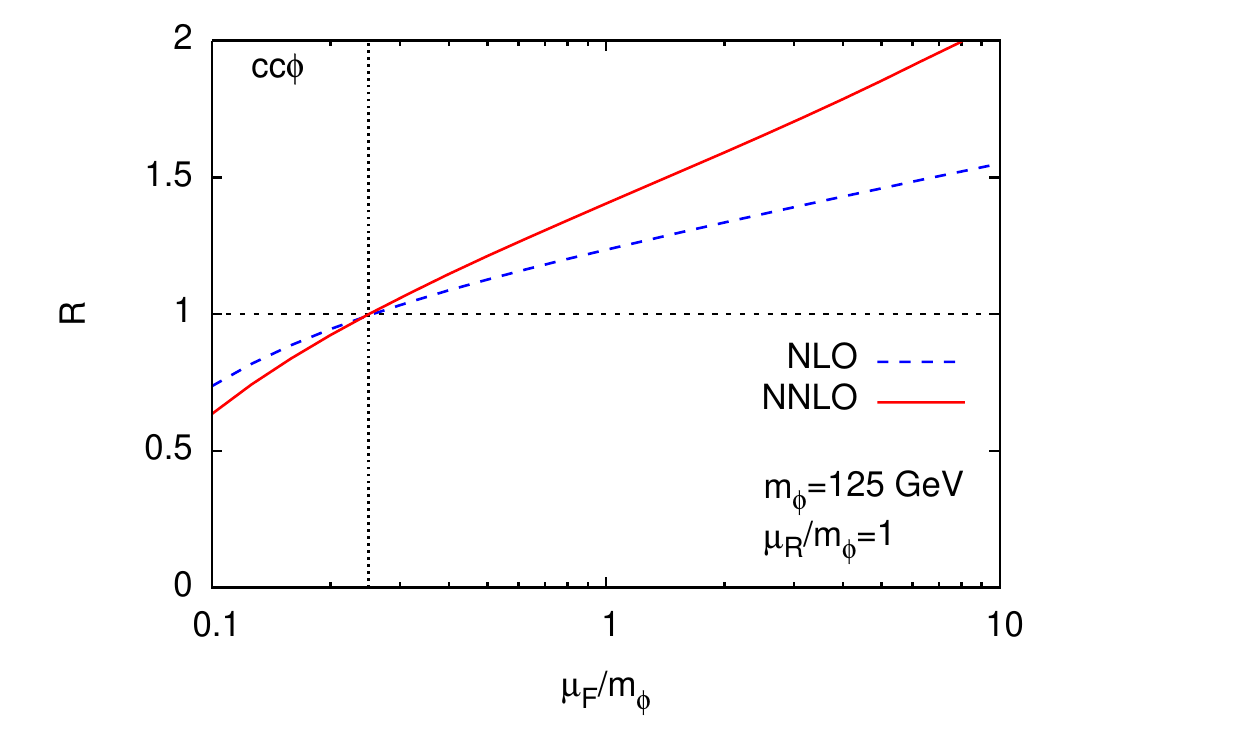}&
      \includegraphics[viewport=20 0 300 220,height=.25\textheight]{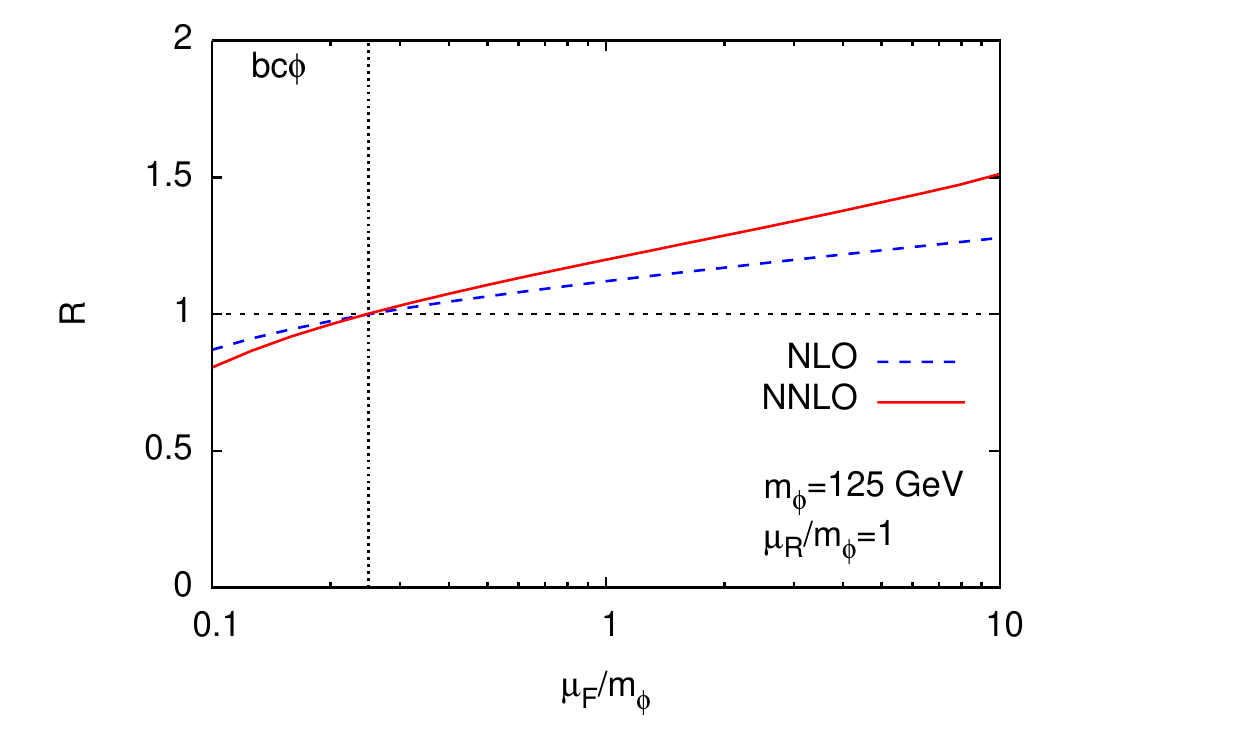}\\[0em]
      \includegraphics[viewport=20 0 300 220,height=.25\textheight]{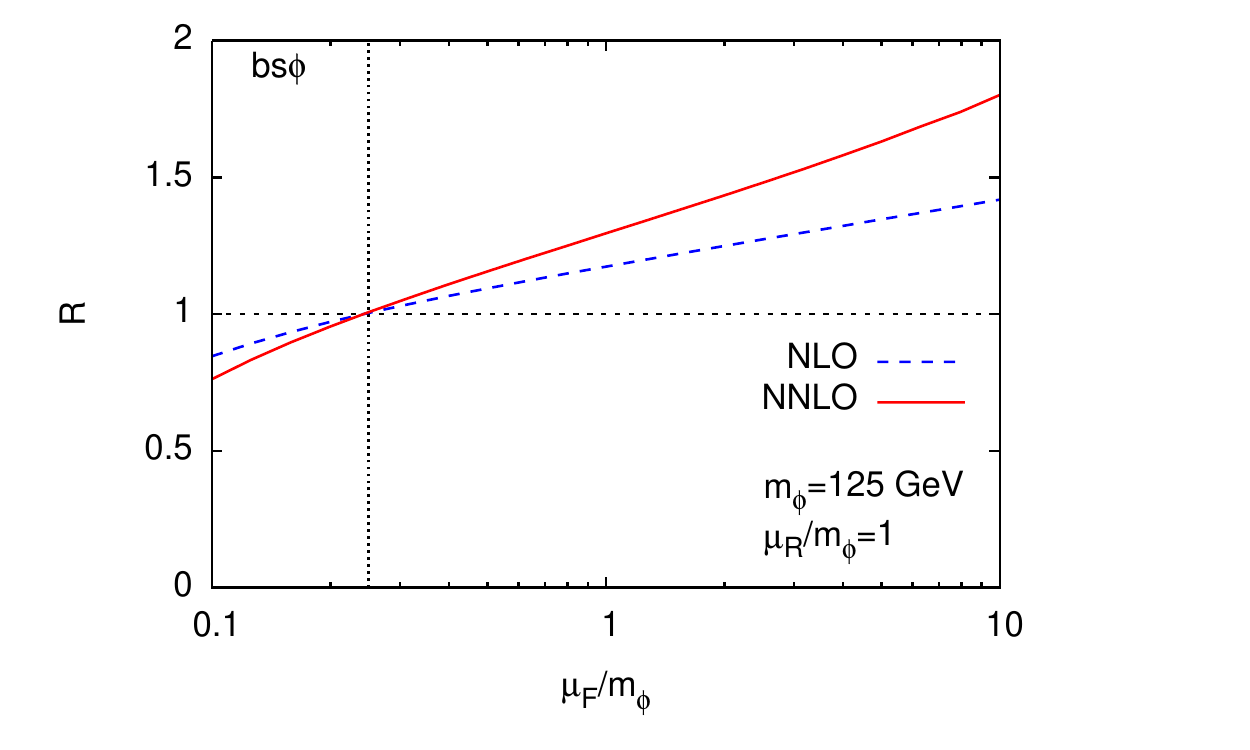}&
      \includegraphics[viewport=20 0 300 220,height=.25\textheight]{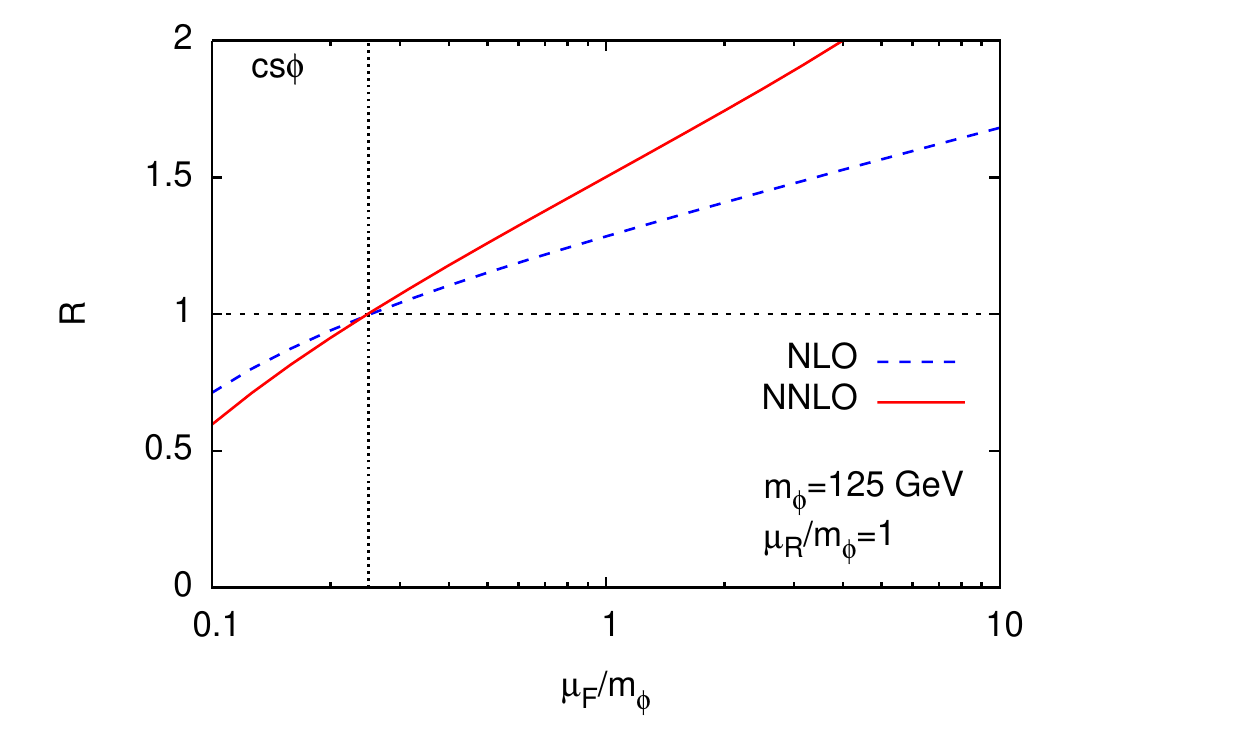}
    \end{tabular}
    \parbox{.9\textwidth}{
      \caption[]{\label{fig:rat-125}\sloppy The double ratio $R_{\qqh}$
        defined in \eqn{eq:rqqh} as a function of the factorization
        scale $\muF$ for $\mphi=125$\,GeV and $\muR=\mphi$.  Upper left
        to lower right: $\qqh=\cch,\bch,\bsh,\csh$.  Dashed: \nlo{};
        Solid: \nnlo{}.}}
  \end{center}
\end{figure}

\begin{figure}
  \begin{center}
    \begin{tabular}{cc}
      \includegraphics[viewport=20 0 300 220,height=.25\textheight]{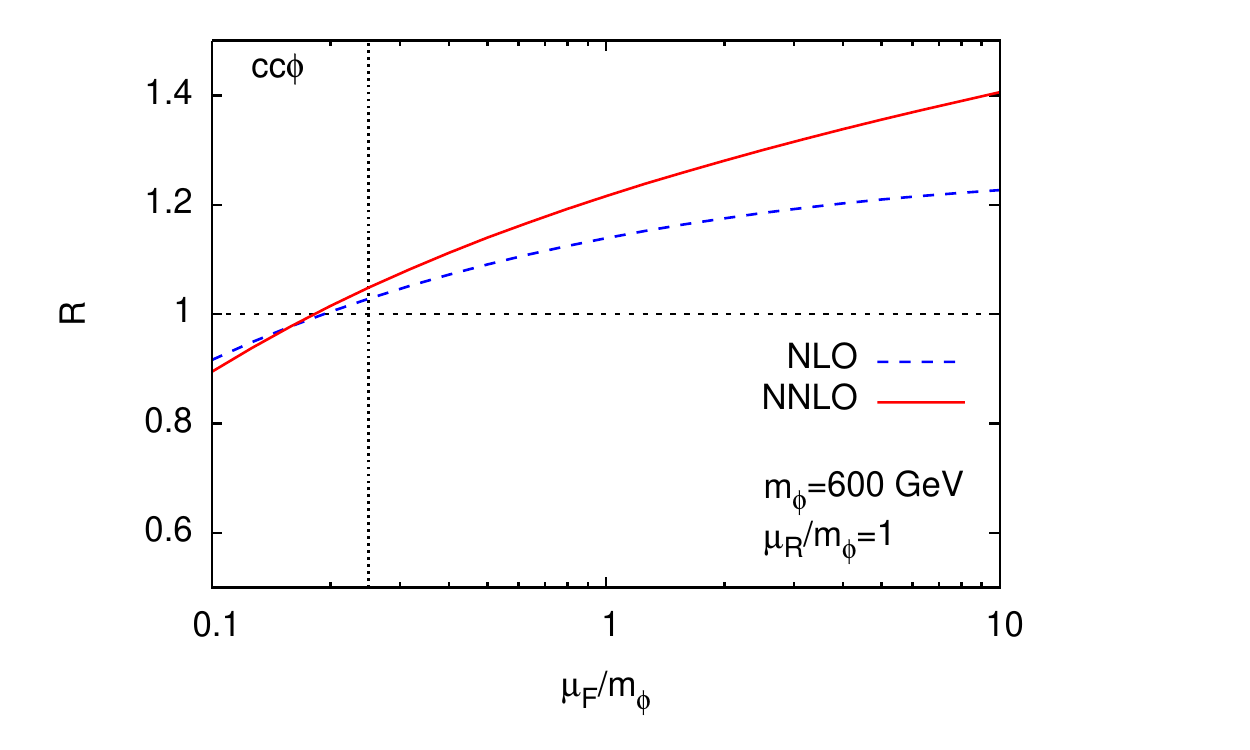}&
      \includegraphics[viewport=20 0 300 220,height=.25\textheight]{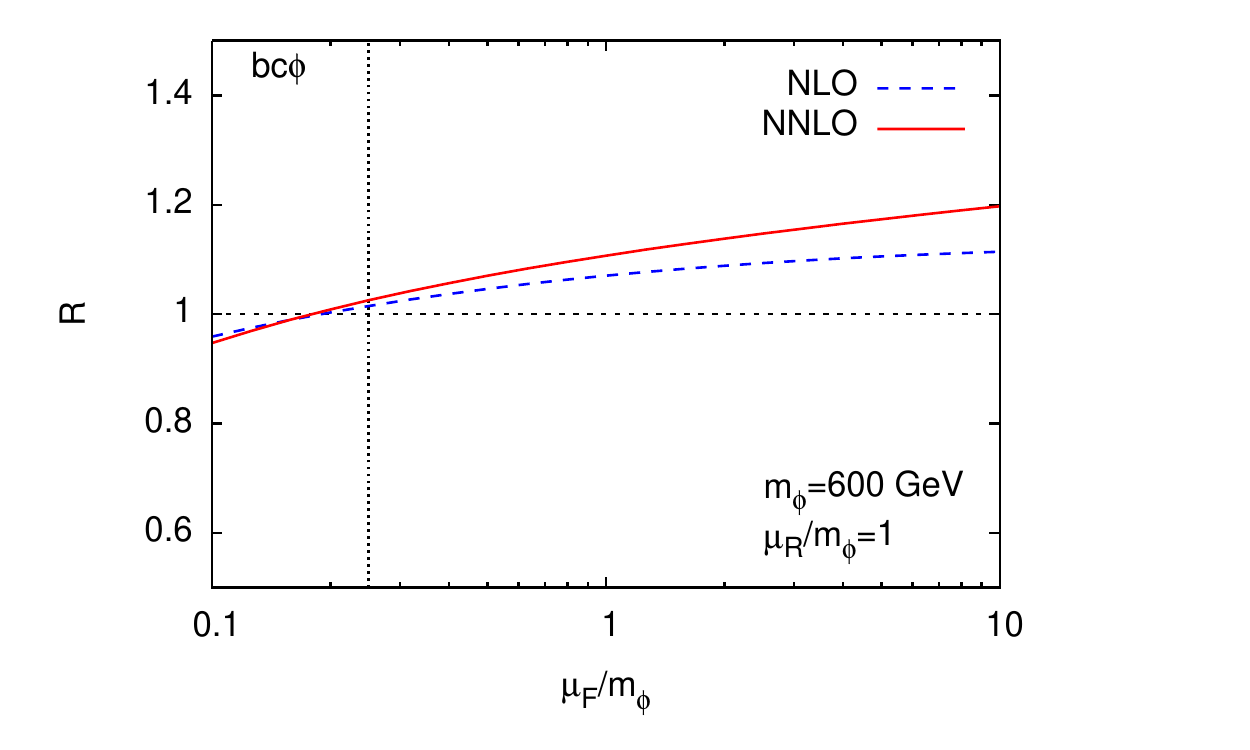}\\
      \includegraphics[viewport=20 0 300 220,height=.25\textheight]{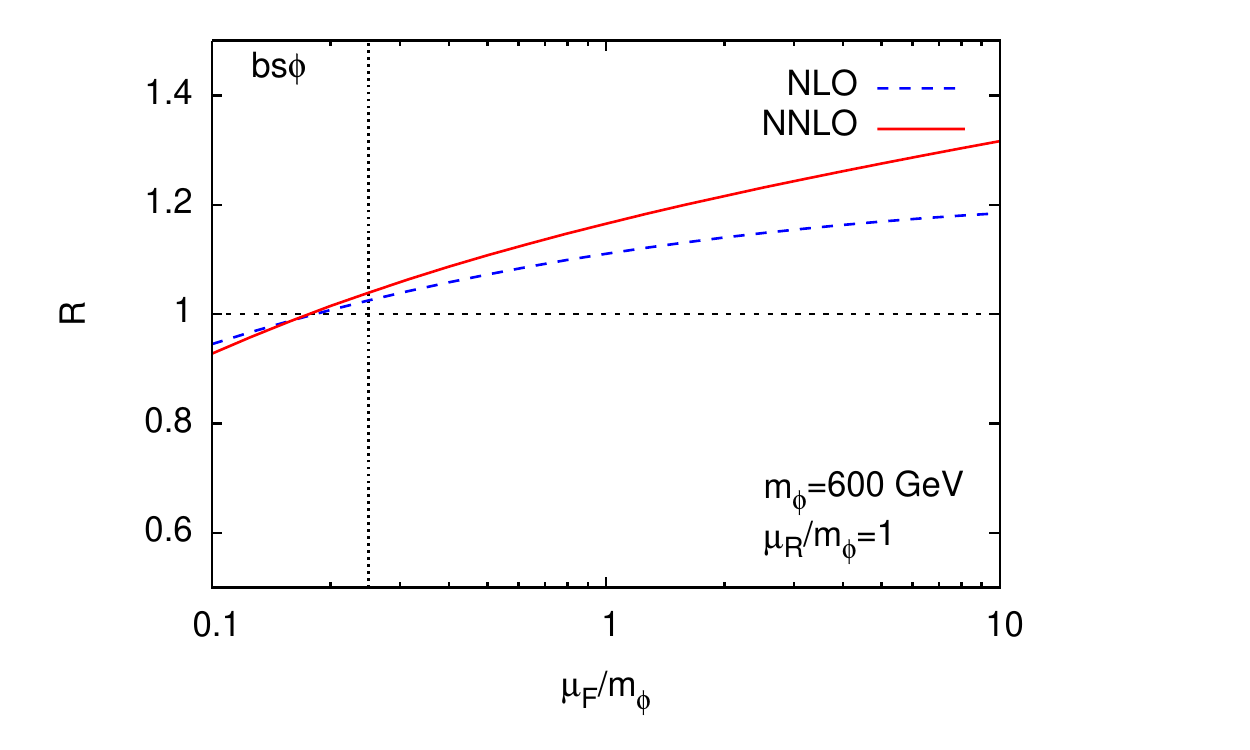}&
      \includegraphics[viewport=20 0 300 220,height=.25\textheight]{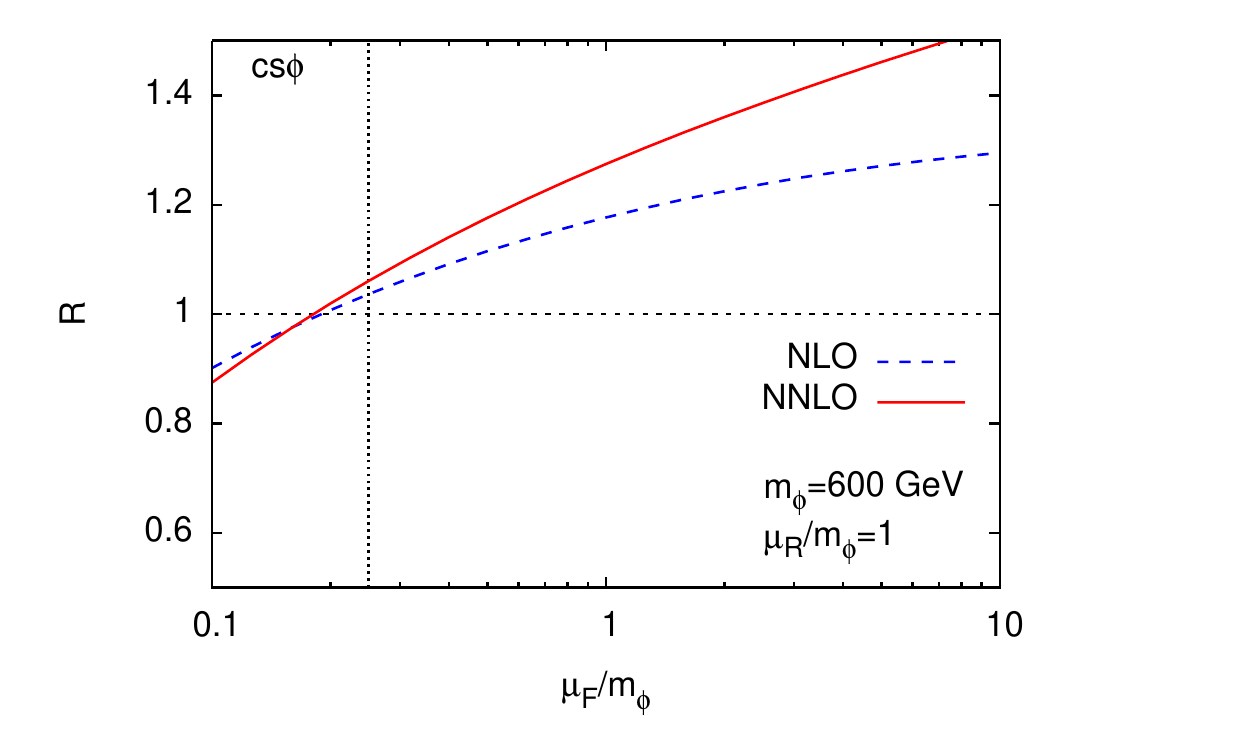}
    \end{tabular}
    \parbox{.9\textwidth}{
      \caption[]{\label{fig:rat-600}\sloppy
        Same as \fig{fig:rat-125}, but for $\mphi=600$\,GeV.
        }}
  \end{center}
\end{figure}

Any deviation from $R_{\qqh}=1$ is due to the hard scattering
process. These effects depend on the choice of $\muRhat$ and $\muFhat$.
\fig{fig:rat-125} shows the $\muFhat$ dependence of $R_{\qqh}$ at
$\muRhat=1$ for $\mphi=125$\,GeV; the effects typically
decrease/increase towards larger/smaller $\muRhat$. Interestingly, one
observes that also here the scale $\muFhat=1/4$ plays a special role: in
all cases, the hard-scattering effects become minimal around this
value. Moreover, we have checked that this observation is virtually
independent of $\muRhat$ (at least within $\muRhat\in[1/10,10]$).  For
$\mphi=600$\,GeV, one observes an analogous behavior, albeit again at
slightly lower $\muFhat$, see \fig{fig:rat-600}.

\subsection{Remarks on the central scale}\label{sec:philo}

Despite the fact that the ``determination'' of a central scale is not an
unambiguous concept, it may be worth recalling the special role of this
quantity for the $\qqh$ processes. Let us focus on the case
$\mphi=125$\,GeV for the sake of simplicity. On the one hand, our
studies of Section\,\ref{sec:scales} have singled out the values
$(\muRhat,\muFhat)=(1,1/4)$ as preferable from a perturbative point of
view, see also \citere{Harlander:2003ai}. As noted above, this is
compatible with kinematic considerations based on the behavior of the
bottom-quark parton
densities\,\cite{Boos:2003yi,Maltoni:2003pn,Rainwater:2002hm}. In
addition, it turns out that, for the $\bbh$ process, the \lo{}
perturbative predictions within the \fs{4} and the \fs{5} agree at the
5-10\% level for this choice of scales, while they differ by more than a
factor of four for $(\muRhat,\muFhat)=(1,1)$, for
example\,\cite{Kramer:2004ie}. Furthermore, it was found that the total
cross section is almost completely exhausted by the partonic $b\bar b$
channel at $(\muRhat,\muFhat)=(1,1/4)$, while all other channels are
very small\,\cite{Harlander:2006pc}. Specifically, for $\mphi=125$\,GeV,
one finds
\begin{equation}
\begin{split}
\sigma(\bbh) = \sigma(\bbh)\left[ 1.04 - 0.06 + 0.02 \right]\quad
\mbox{at}\quad (\muRhat,\muFhat)=(1,1/4)\,,
\end{split}
\end{equation}
where the first number in square brackets denotes the $b\bar b$, the
second the $(b+\bar b)g$, and the third the sum of the remaining
channels.  For $(\muRhat,\muFhat)=(1,1)$, on the other hand, we find a
large cancellation between the $b\bar b$ and the $(b+\bar b)g$ channel:
\begin{equation}
\begin{split}
\sigma(\bbh) = \sigma(\bbh)\left[ 2.39 - 1.56 + 0.17 \right]\quad
\mbox{at}\quad (\muRhat,\muFhat)=(1,1)\,.
\end{split}
\end{equation}
That latter observation seems closely related to the one of
Section\,\ref{sec:hard} which shows that the differences between
different initial state partons due to genuinely hard effects
practically vanish at $(\muRhat,\muFhat)=(1,1/4)$. 
For the sake of completeness, let us note that analogous observations
hold at larger Higgs masses, albeit at slightly different values for the
central scales. 

The author is not aware of any solid theoretical arguments that would
explain why all of these observations occur at this one particular
choice for the unphysical scales; one may simply characterize them as a
mere couriosity. On the other hand, if there is an explanation, it would
certainly be useful for the reduction of the theoretical uncertainty in
perturbative calculations.

\clearpage

\section{Conclusions}\label{sec:conclusions}

Cross sections for the production of neutral and charged scalar and
pseudo-scalar particles via quark annihilation have been calculated
through \nnlo{} \qcd{}. The results should be useful for studying models
with an extended Higgs sector at hadron colliders. Explicit predictions
for cross sections in exemplary cases of a \sm{}-like and a heavy Higgs
boson have been provided. For more general analyses, the next version of
the public program \sushi{}\,\cite{Harlander:2012pb} will provide easy
access to these cross sections.

As a final remark, we note that the considerations of
Section\,\ref{sec:calc} can be extended to higher orders of perturbation
theory, and are not restricted to the total inclusive cross section. The
\nklo{3} results of \citeres{Ahmed:2014cha,Ahmed:2014era} as well as the
differential results of
\citeres{Buehler:2012cu,Harlander:2014hya,Harlander:2011fx,Harlander:2010cz}
can therefore be generalized to $\qqh$ processes by a simple replacement
of the parton densities as well.

\paragraph{Acknowledgments.}
The work was motivated by activities of the {\it LHC Higgs Cross Section
  Working Group}. The author would like to thank Marco Bonvini, Stefan
Liebler, and Frank Tackmann for constructive comments on the manuscript,
and Marisa Sandhoff and Torsten Harenberg for administration of the DFG
FUGG cluster at Bergische Universit\"at Wuppertal, where most of the
calculations for this paper were performed. Financial support by DFG,
contract HA\,2990/6-1 is greatly acknowledged.

\def\app#1#2#3{{\it Act.~Phys.~Pol.~}\jref{\bf B #1}{#2}{#3}}
\def\apa#1#2#3{{\it Act.~Phys.~Austr.~}\jref{\bf#1}{#2}{#3}}
\def\annphys#1#2#3{{\it Ann.~Phys.~}\jref{\bf #1}{#2}{#3}}
\def\cmp#1#2#3{{\it Comm.~Math.~Phys.~}\jref{\bf #1}{#2}{#3}}
\def\cpc#1#2#3{{\it Comp.~Phys.~Commun.~}\jref{\bf #1}{#2}{#3}}
\def\epjc#1#2#3{{\it Eur.\ Phys.\ J.\ }\jref{\bf C #1}{#2}{#3}}
\def\fortp#1#2#3{{\it Fortschr.~Phys.~}\jref{\bf#1}{#2}{#3}}
\def\ijmpc#1#2#3{{\it Int.~J.~Mod.~Phys.~}\jref{\bf C #1}{#2}{#3}}
\def\ijmpa#1#2#3{{\it Int.~J.~Mod.~Phys.~}\jref{\bf A #1}{#2}{#3}}
\def\jcp#1#2#3{{\it J.~Comp.~Phys.~}\jref{\bf #1}{#2}{#3}}
\def\jetp#1#2#3{{\it JETP~Lett.~}\jref{\bf #1}{#2}{#3}}
\def\jphysg#1#2#3{{\small\it J.~Phys.~G~}\jref{\bf #1}{#2}{#3}}
\def\jhep#1#2#3{{\small\it JHEP~}\jref{\bf #1}{#2}{#3}}
\def\mpla#1#2#3{{\it Mod.~Phys.~Lett.~}\jref{\bf A #1}{#2}{#3}}
\def\nima#1#2#3{{\it Nucl.~Inst.~Meth.~}\jref{\bf A #1}{#2}{#3}}
\def\npb#1#2#3{{\it Nucl.~Phys.~}\jref{\bf B #1}{#2}{#3}}
\def\nca#1#2#3{{\it Nuovo~Cim.~}\jref{\bf #1A}{#2}{#3}}
\def\plb#1#2#3{{\it Phys.~Lett.~}\jref{\bf B #1}{#2}{#3}}
\def\prc#1#2#3{{\it Phys.~Reports }\jref{\bf #1}{#2}{#3}}
\def\prd#1#2#3{{\it Phys.~Rev.~}\jref{\bf D #1}{#2}{#3}}
\def\pR#1#2#3{{\it Phys.~Rev.~}\jref{\bf #1}{#2}{#3}}
\def\prl#1#2#3{{\it Phys.~Rev.~Lett.~}\jref{\bf #1}{#2}{#3}}
\def\pr#1#2#3{{\it Phys.~Reports }\jref{\bf #1}{#2}{#3}}
\def\ptp#1#2#3{{\it Prog.~Theor.~Phys.~}\jref{\bf #1}{#2}{#3}}
\def\ppnp#1#2#3{{\it Prog.~Part.~Nucl.~Phys.~}\jref{\bf #1}{#2}{#3}}
\def\rmp#1#2#3{{\it Rev.~Mod.~Phys.~}\jref{\bf #1}{#2}{#3}}
\def\sovnp#1#2#3{{\it Sov.~J.~Nucl.~Phys.~}\jref{\bf #1}{#2}{#3}}
\def\sovus#1#2#3{{\it Sov.~Phys.~Usp.~}\jref{\bf #1}{#2}{#3}}
\def\tmf#1#2#3{{\it Teor.~Mat.~Fiz.~}\jref{\bf #1}{#2}{#3}}
\def\tmp#1#2#3{{\it Theor.~Math.~Phys.~}\jref{\bf #1}{#2}{#3}}
\def\yadfiz#1#2#3{{\it Yad.~Fiz.~}\jref{\bf #1}{#2}{#3}}
\def\zpc#1#2#3{{\it Z.~Phys.~}\jref{\bf C #1}{#2}{#3}}
\def\ibid#1#2#3{{ibid.~}\jref{\bf #1}{#2}{#3}}
\def\otherjournal#1#2#3#4{{\it #1}\jref{\bf #2}{#3}{#4}}
\newcommand{\jref}[3]{{\bf #1}, #3 (#2)}
\newcommand{\hepph}[1]{\href{http://arXiv.org/abs/hep-ph/#1}{{\tt hep-ph/#1}}}
\newcommand{\hepth}[1]{\href{http://arXiv.org/abs/hep-th/#1}{{\tt hep-th/#1}}}
\newcommand{\heplat}[1]{\href{http://arXiv.org/abs/hep-lat/#1}{{\tt hep-lat/#1}}}
\newcommand{\mathph}[1]{\href{http://arXiv.org/abs/math-ph/#1}{{\tt math-ph/#1}}}
\newcommand{\arxiv}[2]{\href{http://arXiv.org/abs/#1}{{\tt arXiv:#1}}}
\newcommand{\bibentry}[4]{#1, {\it #2}, #3\ifthenelse{\equal{#4}{}}{}{,
}#4.}

\end{document}

%% file: xsecs-MMHT2014.tex
\begin{tabular}{|c|l|D{,}{\cdot}{17.4}|}
\hline$ M_\phi/$GeV & $\qqh$ & \multicolumn{1}{c|}{$\sigma/\betaqq/ $pb} \\
\hline\multirow{5}{*}{125} & $b\bar b\phi $  & (  5.23\pm   0.58\pm   0.11), 10^{-1} \\
 & $c\bar c\phi $  & (  1.64\pm   0.08\pm   0.04), 10^{0} \\
 & $b\bar c\phi^+ $  & (  9.35\pm   0.76\pm   0.23), 10^{-1} \\
 & $b\bar s\phi $  & (  1.39\pm   0.09\pm   0.12), 10^{0} \\
 & $c\bar s\phi^- $  & (  2.44\pm   0.10\pm   0.21), 10^{0} \\
\hline\multirow{5}{*}{600} & $b\bar b\phi $  & (  1.19\pm   0.02\pm   0.05), 10^{-3} \\
 & $c\bar c\phi $  & (  3.38\pm   0.02\pm   0.15), 10^{-3} \\
 & $b\bar c\phi^+ $  & (  2.02\pm   0.03\pm   0.08), 10^{-3} \\
 & $b\bar s\phi $  & (  2.97\pm   0.03\pm   0.28), 10^{-3} \\
 & $c\bar s\phi^- $  & (  4.95\pm   0.03\pm   0.46), 10^{-3} \\
\hline\multirow{5}{*}{750} & $b\bar b\phi $  & (  4.04\pm   0.06\pm   0.19), 10^{-4} \\
 & $c\bar c\phi $  & (  1.15\pm   0.01\pm   0.06), 10^{-3} \\
 & $b\bar c\phi^+ $  & (  6.87\pm   0.06\pm   0.34), 10^{-4} \\
 & $b\bar s\phi $  & (  1.01\pm   0.01\pm   0.10), 10^{-3} \\
 & $c\bar s\phi^- $  & (  1.70\pm   0.01\pm   0.17), 10^{-3} \\
\hline\end{tabular}

%% file: cch-arxiv.bbl
\begin{thebibliography}{99}
%
%
%
%

\bibitem{Gunion:1989we}
  \bibentry{J.F.~Gunion, H.E.~Haber, G.L.~Kane, and S.~Dawson}
  {The Higgs Hunter's Guide}
  {Front.\ Phys.\  {\bf 80} (2000) 1}
   {}
%
%

\bibitem{Dittmaier:2011ti}
  \bibentry{The LHC Higgs Cross Section Working
  Group Collaboration}
  {Handbook of LHC Higgs cross sections: 1. Inclusive observables}
  {\arxiv{1101.0593}{hep-ph}}
  {}
%
%

\bibitem{Dittmaier:2012vm}
  \bibentry{The LHC Higgs Cross Section Working
  Group Collaboration} {Handbook of LHC Higgs cross sections:
  2. Differential distributions} {\arxiv{1201.3084}{hep-ph}}{}
%
%

\bibitem{Heinemeyer:2013tqa}
  \bibentry{The LHC Higgs Cross Section Working Group Collaboration}
  {Handbook of LHC Higgs Cross Sections: 3. Higgs Properties}
  {\arxiv{1307.1347}{hep-ph}}
   {}
%
%

\bibitem{Delaunay:2013pja}
  \bibentry{C.~Delaunay, T.~Golling, G.~Perez, and Y.~Soreq}
  {Enhanced Higgs boson coupling to charm pairs}
  {\prd{89}{2014}{033014}}
  {\arxiv{1310.7029}{hep-ph}}
%
%

\bibitem{Gomez:2015duj}
  \bibentry{M.E.~G\'omez, S.~Heinemeyer, and M.~Rehman}
  {The Quark Flavor Violating Higgs Decay 
  $h \rightarrow \bar b s + b \bar s$ in the MSSM}
  {\arxiv{1511.04342}{hep-ph}}
   {}
%

\bibitem{Buttar:2006zd}
\bibentry{C.~Buttar {\it et al.}}
{Les Houches physics at TeV colliders 2005, standard model, QCD, EW, and
Higgs working group: Summary report}
{\hepph{0604120}}
{}
%
%

\bibitem{Harlander:2011aa}
\bibentry{R. Harlander, M. Kr\"amer, M. Schumacher}
{Bottom-quark associated Higgs-boson production: reconciling the four-
and five-flavour scheme approach}
{\arxiv{1112.3478}{hep-ph}}
{\href{https://twiki.cern.ch/twiki/pub/LHCPhysics/MSSMNeutral/santandermatching-hks.pdf}{LHC Higgs Cross Section Working Group Wiki Page}}
%
%

\bibitem{Dittmaier:2003ej}
\bibentry{S.~Dittmaier, M.~Kr\"amer, and M.~Spira}
{Higgs radiation off bottom quarks at the Tevatron and the LHC}
{\prd{70}{2004}{074010}}
{\hepph{0309204}}
%
%

\bibitem{Dawson:2003kb}
\bibentry{S.~Dawson, C.B.~Jackson, L.~Reina, and D.~Wackeroth}
{Exclusive Higgs boson production with bottom quarks at hadron colliders}
{\prd{69}{2004}{074027}}
{\hepph{0311067}}
%
%

\bibitem{Harlander:2003ai} 
\bibentry{R.V.~Harlander and W.B.~Kilgore}
  {Higgs boson production in bottom quark fusion at
     next-to-next-to-leading order}
     {\prd{68}{2003}{013001}}
    {\hepph{0304035}}
%
%

\bibitem{Maltoni:2012pa}
  \bibentry{F.~Maltoni, G.~Ridolfi, and M.~Ubiali}
  {b-initiated processes at the LHC: a reappraisal}
  {\jhep{1207}{2012}{022}}
  {\arxiv{1203.6393}{hep-ph}}
%
%

\bibitem{Forte:2015hba}
  \bibentry{S.~Forte, D.~Napoletano and M.~Ubiali}
  {Higgs production in bottom-quark fusion in a matched scheme}
  {\plb{751}{2015}{331}}
  {\arxiv{1508.01529}{hep-ph}}
%

\bibitem{Bonvini:2015pxa}
  \bibentry{M.~Bonvini, A.S.~Papanastasiou, and F.J.~Tackmann}
  {Resummation and Matching of $b$-quark Mass Effects in $b\bar{b}H$ Production}
  {\arxiv{1508.03288}{hep-ph}}
   {}
%
%

\bibitem{Wiesemann:2014ioa}
  \bibentry{M.~Wiesemann, R.~Frederix, S.~Frixione, V.~Hirschi,
  F.~Maltoni, and P.~Torrielli} {Higgs production in association with
  bottom quarks} {\jhep{1502}{2015}{132}} {\arxiv{1409.5301}{hep-ph}}
%
%

\bibitem{Hamberg:1991np}
\bibentry{R.~Hamberg, W.L.~van Neerven, and T.~Matsuura}
{A complete calculation of the order $\alpha_s^2$ correction to the
Drell-Yan K factor}
{\npb{359}{1991}{343}, (E)~\ibid{B 644}{2002}{403}}
{}
%
%

\bibitem{Kniehl:1990iv}
\bibentry{B.A.~Kniehl}
{Associated production of Higgs and $Z$ bosons from gluon fusion in hadron
collisions}
{\prd{42}{2253}{1990}}
{}
%
%
%
%

\bibitem{Dicus:1988yh}
  \bibentry{D.A.~Dicus and C.~Kao}
  {Higgs Boson-$Z^0$ Production From Gluon Fusion}
  {\prd{38}{1988}{1008}; (E) \ibid{D 42}{1990}{2412}}
  {}
%
%

\bibitem{Brein:2003wg}
\bibentry{O.~Brein, A.~Djouadi, and R.~Harlander}
{{\abbrev NNLO} QCD corrections to the Higgs-strahlung
processes at hadron colliders}
{\plb{579}{2004}{149}}
{\hepph{0307206}}
%
%

\bibitem{Harlander:2012pb}
  \bibentry{R.V.~Harlander, S.~Liebler, and H.~Mantler}
  {SusHi: A program for the calculation of Higgs production in gluon fusion and bottom-quark annihilation in the Standard Model and the MSSM}
  {\cpc{184}{2013}{1605}}
  {\arxiv{1212.3249}{hep-ph}}
%
%

\bibitem{Boos:2003yi}
\bibentry{E.~Boos and T.~Plehn}
{Higgs-boson production induced by bottom quarks}
{\prd{69}{2004}{094005}}
{\hepph{0304034}}
%
%

\bibitem{Maltoni:2003pn}
\bibentry{F.~Maltoni, Z.~Sullivan, and S.~Willenbrock}
{Higgs-boson production via bottom-quark fusion}
{\prd{67}{2003}{093005}}
{\hepph{0301033}}
%
%

\bibitem{Rainwater:2002hm}
\bibentry{D.~Rainwater, M.~Spira, and D.~Zeppenfeld}
{Higgs boson production at hadron colliders: Signal and background processes}
{\hepph{0203187}}
{}
%
%

\bibitem{Chetyrkin:2000yt}
  \bibentry{K.G.~Chetyrkin, J.H.~K\"uhn, and M.~Steinhauser}
  {RunDec: a Mathematica package for running and decoupling of the
  strong coupling and quark masses}
  {\cpc{133}{2000}{43}}
  {\hepph{0004189}}
%
%

\bibitem{Martin:2009iq}
  \bibentry{A.D.~Martin, W.J.~Stirling, R.S.~Thorne, and G.~Watt}
  {Parton distributions for the LHC}
  {\epjc{63}{2009}{189}}
  {\arxiv{0901.0002}{hep-ph}}
%
%

\bibitem{Buckley:2014ana}
  \bibentry{A.~Buckley, J.~Ferrando, S.~Lloyd, K.~Nordstr\"om, B.~Page,
  M.~R\"ufenacht, M.~Sch\"onherr, and G.~Watt} {LHAPDF6: parton density
  access in the LHC precision era} {\epjc{75}{2015}{3}}
  {\arxiv{1412.7420}{hep-ph}}
%
%

\bibitem{lhapdf}
 \bibentry{M.R.~Whalley, D.~Bourilkov, and R.C.~Group}
  {The Les Houches accord PDFs (LHAPDF) and LHAGLUE}
  {\hepph{0508110}}
  {\href{http://projects.hepforge.org/lhapdf/}
  {\tt http://projects.hepforge.org/lhapdf/}}
%
%

\bibitem{Harland-Lang:2014zoa}
  \bibentry{L.A.~Harland-Lang, A.D.~Martin, P.~Motylinski, and R.S.~Thorne}
  {Parton distributions in the LHC era: MMHT 2014 PDFs}
  {\epjc{75}{2015}{204}}
  {\arxiv{1412.3989}{hep-ph}}
%
%

\bibitem{Dulat:2015mca}
  \bibentry{S.~Dulat {\it et al.}}
  {The CT14 Global Analysis of Quantum Chromodynamics}
  {\arxiv{1506.07443}{hep-ph}}
   {}
%
%

\bibitem{Ball:2014uwa}
  \bibentry{R.D.~Ball {\it et al.} [NNPDF Collaboration]}
  {Parton distributions for the LHC Run II}
  {\jhep{1504}{2015}{040}}
  {\arxiv{1410.8849}{hep-ph}}
%
%

\bibitem{Bagnaschi:2014zla}
  \bibentry{E.~Bagnaschi, R.V.~Harlander, S.~Liebler, H.~Mantler, 
P.~Slavich, and A.~Vicini}
  {Towards precise predictions for Higgs-boson production in the MSSM}
   {\jhep{1406}{2014}{167}}
  {\arxiv{1404.0327}{hep-ph}}
%
%

\bibitem{Kramer:2004ie}
\bibentry{M.~Kr\"amer}
{Associated Higgs production with bottom quarks at hadron colliders}
{\hepph{0407080}}
{}
%
%

\bibitem{Harlander:2006pc}
\bibentry{R.~Harlander}
{Standard and SUSY Higgs production at the LHC}
{Pramana {\bf 67} (2006) 875 }
{\hepph{0606095}}
%
%

\bibitem{Ahmed:2014cha}
  \bibentry{T.~Ahmed, N.~Rana, and V.~Ravindran}
  {Higgs boson production through 
  $b \bar b$ annihilation at threshold in N$^3$LO QCD}
  {\jhep{1410}{2014}{139}}
  {\arxiv{1408.0787}{hep-ph}}
%
%

\bibitem{Ahmed:2014era}
  \bibentry{T.~Ahmed, M.K.~Mandal, N.~Rana, and V.~Ravindran}
  {Higgs Rapidity Distribution in $b {\bar b}$ Annihilation at Threshold in N$^{3}$LO QCD}
  {\jhep{1502}{2015}{131}}
  {\arxiv{1411.5301}{hep-ph}}
%
%

\bibitem{Buehler:2012cu}
  \bibentry{S.~Buehler, F.~Herzog, A.~Lazopoulos, and R.~Mueller}
  {The Fully differential hadronic production of a Higgs boson via bottom quark fusion at NNLO}
  {\jhep{1207}{2012}{115}}
  {\arxiv{1204.4415}{hep-ph}}
%
%

\bibitem{Harlander:2014hya}
  \bibentry{R.V.~Harlander, A.~Tripathi, and M.~Wiesemann} {Higgs
  production in bottom quark annihilation: Transverse momentum
  distribution at NNLO+NNLL} {\prd{90}{2014}{015017}}
  {\arxiv{1403.7196}{hep-ph}}
%
%

\bibitem{Harlander:2011fx}
  \bibentry{R.~Harlander and M.~Wiesemann}
  {Jet-veto in bottom-quark induced Higgs production at next-to-next-to-leading order}
  {\jhep{1204}{2012}{066}}
  {\arxiv{1111.2182}{hep-ph}}
%
%

\bibitem{Harlander:2010cz}
\bibentry{R.V.~Harlander, K.J.~Ozeren, and M.~Wiesemann}
{Higgs plus jet production in bottom quark annihilation at
  next-to-leading order}
{\plb{693}{2010}{269}}
{\arxiv{1007.5411}{hep-ph}}
%
%


\end{thebibliography}
